\documentclass[a4paper,11pt]{article}
\usepackage{jheppub} % for details on the use of the package, please
\pdfoutput=1 
\usepackage{graphicx,dcolumn,bm,epsfig,cancel,hyperref}
\usepackage{multirow}
\usepackage{amsmath}
\usepackage{amssymb, times}
\newcommand{\comment}[1]{{}}
\usepackage{color}

\newcommand{\mha}[0]{m_{h_1}}
\newcommand{\mhb}[0]{m_{h_2}}

\newcommand{\blue}[1]{\textcolor{blue}{#1}}

\title{
Collider and Gravitational Wave Complementarity in Exploring the Singlet Extension of the Standard Model
}
\author[a]{Alexandre Alves}
\author[b]{Tathagata Ghosh}
\author[c]{Huai-Ke Guo}
\author[c]{Kuver Sinha}
\author[c]{Daniel Vagie}
\affiliation[a]{Departamento de F\'isica, Universidade Federal de S\~ao Paulo, UNIFESP, Diadema, Brazil}
\affiliation[b]{Department of Physics \& Astronomy, University of Hawaii, Honolulu, HI 96822, USA}
\affiliation[c]{Department of Physics and Astronomy, University of Oklahoma, Norman, OK 73019, USA}
\emailAdd{aalves@unifesp.br}
\emailAdd{tghosh@hawaii.edu}
\emailAdd{ghk@ou.edu}
\emailAdd{kuver.sinha@ou.edu}
\emailAdd{Daniel.d.vagie-1@ou.edu}
\begin{document}

%\date{\today}

\abstract
{
We present a dedicated complementarity study of  gravitational wave and collider measurements of the simplest extension of the Higgs sector: the  singlet scalar augmented Standard Model. We study the following issues: $(i)$ the electroweak phase transition patterns admitted by the model, and the proportion of parameter space for each pattern; $(ii)$ the regions of parameter space that give detectable gravitational waves at future space-based detectors; and $(iii)$ the current and future collider measurements of di-Higgs production, as well as searches for a heavy weak diboson resonance, and how these searches interplay with regions of parameter space that exhibit strong gravitational wave signals. We carefully investigate the behavior of the normalized energy released during the phase transition as a function of the model parameters, address subtle issues pertaining to the bubble wall velocity, and provide a description of different fluid velocity profiles. On the collider side, we identify the subset of points that are most promising in terms of di-Higgs and weak diboson production studies while also giving detectable signals at LISA, setting the stage for future benchmark points that can be used by both communities. 
}

%\pacs{11.30.Er, 11.30.Fs, 11.30.Hv, 12.60.Fr, 31.30.jp}

\maketitle

\section{ Introduction}

Since the first direct detection of gravitational waves (GWs) by the LIGO and Virgo 
collaborations~\cite{Abbott:2016blz}, a new interface has arrived in particle physics -- its intersection with GW astronomy. While ground based GW detectors have their best sensitivity at frequencies $\sim \mathcal{O}(100) \, \text{Hertz}$ and their main targets are black hole and neutron star binaries, there is now growing interest in building space-based interferometer detectors for milli-Hertz or deci-Hertz frequencies. Many detectors 
have been proposed, such as the Laser Interferometer Space Antenna (LISA)~\cite{Audley:2017drz}, the 
Big Bang Observer (BBO), the DECi-hertz Interferometer Gravitational wave 
Observatory (DECIGO)~\cite{Yagi:2011wg}, Taiji~\cite{Gong:2014mca} and Tianqin~\cite{Luo:2015ght}. 
The physical sources of GWs in this
frequency band include supermassive black hole binaries~\cite{Klein:2015hvg}, 
extreme mass ratio inspirals~\cite{Babak:2017tow} and 
the stochastic background of primordial GWs produced during first order cosmological phase 
transitions~\cite{Caprini:2015zlo}.

This offers tremendous opportunities for theorists, as a new window to the early Universe opens up. Aspects of dark sector physics and baryon asymmetry can now be framed fruitfully in a language that lends itself to data from the GW frontier. The key connection is \textit{phase transitions}, which on the one hand are a primary target of future GW experiments, and on the other are important features of scalar potentials and hence have historically been the target of collider physics.

The  purpose of our work is to \textit{explore the complementarity of future GW detectors and future particle colliders in probing phase transitions in the early Universe} -- in the simplest particle physics setting possible, but also with great attention to details within such a setting. The natural choice is the electroweak phase transition (EWPT) ~\cite{Grojean:2006bp} with the simplest extension of the Higgs sector: the  singlet scalar augmented Standard Model or the xSM\footnote{Hidden sector phase transitions are also being actively 
investigated~\cite{Schwaller:2015tja,Alanne:2014bra,Breitbach:2018ddu,Aoki:2017aws,Bai:2018dxf}, and exploring complementarity in such settings is an interesting future direction. We refer to  Ref.~\cite{Caprini:2015zlo,Cai:2017cbj,Weir:2017wfa,Caprini:2018mtu,Mazumdar:2018dfl} for recent work on these topics.}. 
This model is capable of providing a strongly first order EWPT through a tree level barrier and is the simplest model in Class IIA of the tree level renormalizable operators described in~\cite{Chung:2012vg}
(see Ref.~\cite{Beniwal:2018ygo,Croon:2018kqn,Hashino:2018wee,Marzo:2018nov,Beniwal:2017eik,Hashino:2018zsi,Addazi:2017gpt,Addazi:2017nmg,Chiang:2017zbz,Wan:2018udw,Chen:2017cyc,Vieu:2018nfq,Basler:2017uxn,Jinno:2017ixd,Chao:2017ilw,Huang:2014ifa,Tsumura:2017knk,Bian:2017wfv,Hektor:2018esx,Huang:2017rzf,Ghorbani:2017lyk,Kang:2017mkl,Addazi:2017oge,Chen:2017qcz,Marzola:2017jzl,Kobakhidze:2017mru,Zhou:2018zli,Beniwal:2018hyi,Dev:2016feu,Balazs:2016tbi,Ahriche:2018rao,Shajiee:2018jdq,Bian:2018bxr,Blinov:2015sna,Inoue:2015pza,Vaskonen:2016yiu,No:2018fev,Chiang:2018gsn,Chala:2018opy,Cai:2017tmh,Alanne:2018brf,Kannike:2019wsn,Ashoorioon:2009nf,Cheng:2018ajh,Bi:2015qva} for related studies on EWPT and GW). It has been extensively investigated in phenomenological 
studies~\cite{Profumo:2007wc,Profumo:2014opa,Huang:2017jws,Robens:2015gla}, studies of 
EWPT~\cite{Profumo:2014opa,Profumo:2007wc,Kotwal:2016tex,Espinosa:2011ax,Kozaczuk:2015owa} and di-Higgs 
analyses~\cite{Lewis:2017dme} guided by the requirements of EWPT~\cite{Huang:2017jws}, and electroweak baryogenesis (EWBG).

We perform a detailed scan of this model, shedding light on the following issues: $(i)$ the EWPT patterns admitted by the model, and the proportion of parameter space for each pattern; $(ii)$ the regions of parameter space that give detectable GWs at future space-based detectors; $(iii)$ the current and future collider measurements of di-Higgs production, as well as searches for a heavy weak diboson resonance, and how these searches interplay with regions of parameter space that exhibit strong GW signals; and $(iv)$ the complementarity of collider and GW searches in probing this model.

We first carefully work out and incorporate all phenomenological constraints: boundedness of the Higgs potential from below, electroweak vacuum stability at zero temperature, perturbativity, perturbative unitarity, Higgs signal strength measurements and electroweak precision observables. Then, we identify the regions of parameter space which give large signal-to-noise-ratio (SNR) at LISA. We carefully address subtle issues pertaining to the bubble wall velocity $v_w$, making a distinction between $v_w$, which enters GW calculations, and the velocity $v_+$ that is used in EWBG calculations. The relation between these two velocities is determined from a hydrodynamic analysis by solving the velocity profile surrounding the bubble wall. We provide a description of different fluid velocity profiles and investigate the behavior of the normalized energy released during the phase transition, $\alpha$, which primarily determines the SNR, as a function of the model parameters. On the collider side, we identify the subset of points with large SNR at LISA that are most promising in terms of di-Higgs and weak diboson production studies, setting the stage for future benchmark points. 

Much remains to be understood about the Higgs sector. On the collider side, measuring the Higgs cubic and quartic couplings through double or triple Higgs production, both non-resonant as well as resonant, is an extremely difficult  but central goal of future experiments (see e.g.,~\cite{Azatov:2015oxa, Alves:2017ued, DiVita:2017vrr,Plehn:2005nk,Binoth:2006ym,Bizon:2018syu,Liu:2018peg}). While any deviation of the shape of the Higgs potential from what is expected within the Standard Model (SM) would hint to new physics, the sensitivities of such collider studies are found to be rather low.  The detection of GWs from EWPT in future experiments can offer a complementary method of probing the currently largely unknown Higgs potential. Our work is a step in that direction.

The paper is structured as follows. In Sec.~\ref{sec:model}, we define the Higgs potential and set the notations. The standard phenomenological
analysis is discussed in the following Sec.~\ref{sec:pheno}. The next Sec.~\ref{sec:ewpt} discuss the details of the EWPT and GW calculations, after
which the results and discussions from the full scan is presented in Sec.~\ref{sec:result} and we summarize in Sec.~\ref{sec:summary}.

\section{\label{sec:model}The Model}
In this section, we fix our notation by defining the potential for the gauge singlet extended SM, 
known as the``xSM''. This model is defined with the following potential setup~\cite{Profumo:2007wc,Profumo:2014opa,Huang:2017jws}:
\begin{eqnarray}
  V(H,S) &=& -\mu^2 H^{\dagger} H + \lambda (H^{\dagger}H)^2 
  + \frac{a_1}{2} H^{\dagger} H S  \nonumber \\
  &&  + \frac{a_2}{2} H^{\dagger} H S^2 + \frac{b_2}{2} S^2 + \frac{b_3}{3} S^{3} + \frac{b_4}{4}S^4 ,
  \label{eq:v}
\end{eqnarray} 
where $H^{\text{T}} = (G^+, (v_{\text{EW}} + h + i G^0)/\sqrt{2})$ is the SM Higgs doublet 
and $S=v_s + s$ the real scalar gauge singlet.  All the model parameters in the above equation are real.
The parameters $\mu$ and $b_2$ can be solved from the two minimization conditions around the EW vacuum($\equiv (v_{\text{EW}},v_s)$),
\begin{eqnarray}
&&  \mu^2 = \lambda v_{\text{EW}}^2 + \frac{1}{2} v_s (a_1 + a_2 v_s), \nonumber \\
&&   b_2  =  - \frac{1}{4v_s} [v_{\text{EW}}^2(a_1+2a_2 v_s) + 4v_s^2 (b_3+b_4 v_s)] ,
\end{eqnarray}
and $\lambda, a_1, a_2$ can be replaced by physical parameters $\theta$, $m_{h_1}$ and $m_{h_2}$ from 
the mass matrix diagonalization~\footnote{Here $s_{\theta} \equiv \sin \theta$ and $c_{\theta} \equiv \cos \theta$.}:
\begin{eqnarray}
  &&  \lambda = \frac{\mha^2 c_{\theta}^2 + \mhb^2 s_{\theta}^2}{2 v_{\text{EW}}^2} ,\nonumber \\
  &&  a_1 = \frac{2 v_s}{v^2_{\text{EW}}}[2 v_s^2(2 b_4 + \tilde{b}_3) - \mha^2 - \mhb^2 + c_{2\theta}(\mha^2 - \mhb^2)] , \nonumber \\
  &&  a_2 = \frac{-1}{2 v_{\text{EW}}^2 v_s}[-2 v_s(\mha^2+\mhb^2 - 4 b_4 v_s^2) \nonumber \\
  && \hspace{2cm}+ (\mha^2 - \mhb^2)(2c_{2\theta} v_s - v_{\text{EW}} s_{2\theta}) + 4 \tilde{b}_3 v_s^3] ,
\end{eqnarray}
where $\tilde{b}_3 \equiv b_3/v_s$ and we have defined the physical fields $h_1$ and $h_2$ as
\begin{eqnarray}
h_1 = c_{\theta} h + s_{\theta} s, \quad \quad
h_2 =-s_{\theta} h + c_{\theta} s,  
  \label{}
\end{eqnarray}
with a mixing angle $\theta$. We note that $h_1$ is identified as the SM Higgs while $h_2$ is a heavier scalar.
The coupling of $h_1$ with the SM particles is reduced by a factor of $c_{\theta}$ while the coupling of
$h_2$ with SM particles is $(-s_{\theta})$ times the corresponding SM couplings and vanishes in the case of 
zero mixing angle.

With choices of parameter transformations described above, the potential is fully specified by the following 
five parameters:
\begin{eqnarray}
\centering
v_s, \quad \quad \mhb, \quad \quad \theta, \quad \quad b_3, \quad \quad b_4 .
\end{eqnarray}
The model defined here has several variants in the literature. For example, since the potential can be defined with
a translation in the $S$ direction $S \rightarrow S^{\prime} = S - v_s$, such that $\langle S \rangle=0$, 
the resulting 
potential will take the same form as Eq.~\ref{eq:v} but with the addition of a non-zero tadpole term 
$b_1 S$~\cite{Lewis:2017dme}. The potential and physics remain the same but the parameters in the potential will 
transform accordingly. The transformation rules to and from this basis are given in Appendix ~\ref{sec:tadpole}. 
There is also a variant where there is a spontaneously broken $Z_2$ symmetry $S\rightarrow -S$; this corresponds to a 
subset of the parameter space here where $a_1=b_3=0$.

We further note that we do not include CP-violation in this study since the magnitude of the CP-violation is 
typically very constrained by current electric dipole moment 
searches (e.g.,~\cite{Inoue:2014nva,Chen:2017com,Bian:2017wfv} or the included 
CP-violation may be large but has little effect on EWPT~\cite{Guo:2016ixx}). 

\section{\label{sec:pheno}Phenomenological Constraints}
In this section, we briefly discuss the phenomenological constraints used in our analysis, following
the standard treatments given in Refs.~\cite{Pruna:2013bma,Robens:2015gla,Lewis:2017dme}. The phenomenological discussion includes boundedness of the Higgs potential from below, EW vacuum stability at zero temperature, perturbativity, perturbative unitarity,
Higgs signal strength measurements and electroweak precision observables.

First, the potential needs to be bounded from below. Requiring this for arbitrary field directions gives us the 
condition~\cite{Lewis:2017dme}~\footnote{
Note that these tree level relations will change when loop corrections are taken into account.
However due to the way of calculating the effective potential in Eq.(4.1), these relations
suffice to guarantee that the potential is bounded from below when $T \rightarrow 0$ in 
Eq.~\ref{eq:vt}.
},
\begin{equation}
\lambda > 0, \quad b_4 > 0, \quad a_2 \geqslant -2 \sqrt{\lambda b_4} .
\end{equation}
Next, the EW vaccum also needs to be stable at zero temperature. 
Using physical parameters as input will automatically guarantee that the EW vacuum is a minimum. To ensure that the above EW vacuum is stable, one should require that no deeper minimum exists in the potential.
%To further 
%require that the EW vacuum is stable, no deeper minimum in any other location should be allowed. 
In our analysis, we find all the minima by firstly solving $\partial V/\partial \phi_i=0$($\phi_1\equiv h$, $\phi_2 \equiv s$)
and subsequently calculating eigenvalues of the Hessian matrix $\{\partial^2 V/\partial \phi_i \partial\phi_j\}$ to determine the nature of the extrema
for each set of parameter input. 

Next, Higgs signal strength measurements in various channels require the couplings of $h_1$ to be not far 
from the SM Higgs couplings. In the xSM, the couplings of $h_1$ to SM particles are reduced by a factor 
of $\cos\theta$, therefore the Higgs signal 
strength is given by $\mu_H = \cos^2\theta$. Experimentally, the most recent ATLAS and CMS combined fit of 
this value is $\mu_H = 1.09^{+0.11}_{-0.10}$~\cite{Khachatryan:2016vau} and
a $\chi^2$ analysis shows that $|\sin \theta|>0.33$ are excluded at $95\%$ CL~\cite{Carena:2018vpt}.

Moreover, unitarity puts constraints on the high energy behavior of particle scatterings. 
Requiring further the perturbativity 
of these scatterings at high energy will lead to constraints on the model. This tree level perturbativity 
requirement is quantified as the condition that the partial wave amplitude $a_l(s)$ for all $2 \rightarrow 2$ 
processes satisfies $|\text{Re} \, a_l(s)|\lesssim 1/2$ for $\sqrt{s}\rightarrow \infty$.
We consider all channels of scalar/vector boson $2 \rightarrow 2$ scatterings at the leading order in the high 
energy expansion, with details of the S-matrix given in Appendix.~\ref{sec:unitarity}.

Electroweak precision measurements, which mainly 
include the $W$ boson mass measurement~\cite{Lopez-Val:2014jva} and the oblique EW corrections~\cite{Peskin:1991sw,Hagiwara:1994pw},  put further constraints on the model.
The $W$ boson mass $m_W$ can be calculated given experimentally measured values of 
$G_F$, $m_Z$ and the fine structure constant at zero momentum transfer $\alpha(0)$~\cite{Lopez-Val:2014jva}. 
The function relating $m_W$ and these three parameters depends on the loop corrections of the vector boson 
self-energies. Comparing this calculated $m_W$ with the experimental measurement 
$m_W^{\text{exp}} = 80.385 \pm 0.015 \text{GeV}$~\cite{Alcaraz:2006mx,Aaltonen:2012bp,D0:2013jba} 
highly constrains the modification of the loop corrections 
by new physics effects. In this model, the modified loop corrections result from reduced Higgs couplings and 
from the presence
of the heavier scalar $h_2$ and are only dependent on $(\theta, m_{h_2})$ at one-loop level. The same parameter
dependence enters the oblique $S, T, U$ parameters and it turns out that the $W$-mass constraint is much more stringent
than that from the oblique corrections~\cite{Lopez-Val:2014jva,Robens:2015gla}. 

%%%%%%%%%%%%%%%%%%%%%%%%%%%%%%%%%%%%%%%%%%%%%%%%%%%%%%%%%%%%%%%%%%%%%%%%%%%%%%%%%%%%%%%%%%%%%%%%%%%%%%%%%%%%%%%%%
%%%%%%%%%%%%%%%%%%%%%%%%%%%%%%%%%%%%%%%%%%%%%%%%%%%%%%%%%%%%%%%%%%%%%%%%%%%%%%%%%%%%%%%%%%%%%%%%%%%%%%%%%%%%%%%%%
\begin{figure}[t]
\centering
%\includegraphics[width=0.5\textwidth]{fig/a2-b3-SNR-500.eps}
%\quad
\includegraphics[width=0.6\textwidth]{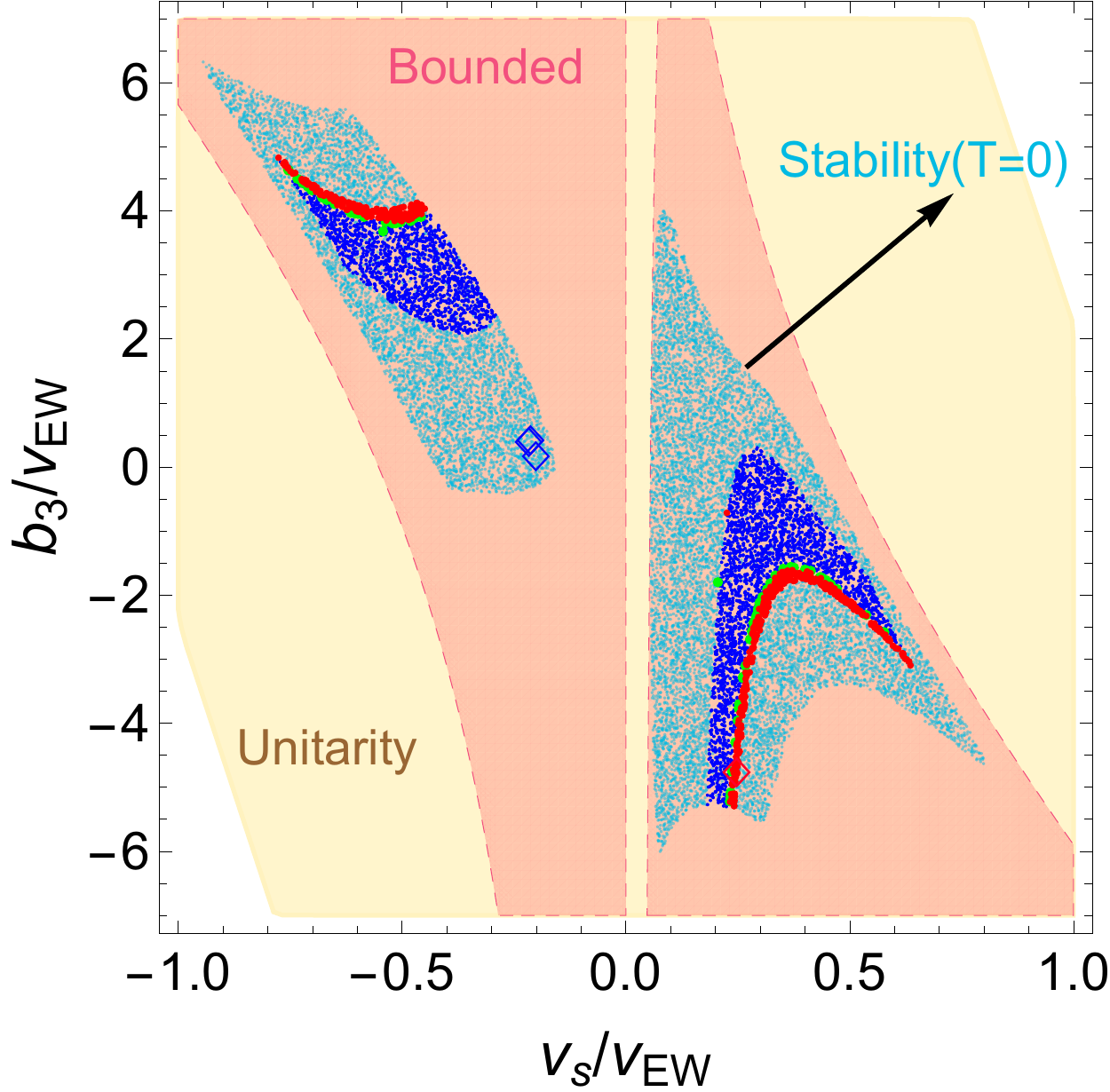}
\caption{
\label{fig:demo}
An illustrative plot showing various phenomenological constraints.
The shaded regions are allowed by requirements of unitarity, boundedness of the potential from below, 
and stability of EW vacuum at zero temperature. Points are also overlapped on this plot where 
various EWPT criteria are fulfilled and with $\text{SNR}>50$ (red), $50>\text{SNR}>10$ (green) and $\text{SNR}<10$ (blue).
The diamond-shaped points give two-step EWPT.
}
\end{figure}
%%%%%%%%%%%%%%%%%%%%%%%%%%%%%%%%%%%%%%%%%%%%%%%%%%%%%%%%%%%%%%%%%%%%%%%%%%%%%%%%%%%%%%%%%%%%%%%%%%%%%%%%%%%%%%%%%
%%%%%%%%%%%%%%%%%%%%%%%%%%%%%%%%%%%%%%%%%%%%%%%%%%%%%%%%%%%%%%%%%%%%%%%%%%%%%%%%%%%%%%%%%%%%%%%%%%%%%%%%%%%%%%%%%

To give the reader a flavor of the above phenomenological constraints, 
we fix $m_{h_2}=300 \, \text{GeV}$, $\theta=0.2$, $b_4=4$ 
and show the various bounds on the remaining two parameters $(v_s/v_{\text{EW}}, b_3/v_{\text{EW}})$ 
in Fig.~\ref{fig:demo}. This choice of $m_{h_2}$ and $\theta$ evades the constraints from the $W$-mass 
as well as the oblique EW corrections and regions outside
the color-shaded regions are excluded by the remaining constraints.
It can be seen from this figure that the least constraining condition comes from the perturbative unitarity requirement 
for this parameter choice. The bounded-from-below condition is more restrictive and also separates the plane 
into two disconnected regions while the stability of the EW vacuum at  zero temperature 
shrinks the allowed parameter space even more.
We also overlaid on this plot the points which pass the various EWPT requirements and give GW signals 
with varying SNR. More details are given in the caption and in the following section.

\section{\label{sec:ewpt}EWPT and Gravitational Waves}

\subsection{Effective Potential}

EWPT is an essential step~\footnote{
Other mechanisms generally do not need EWPT to generate the baryon asymmetry.
For example, in leptogenesis, 
the out-of-equilibrium requirement is provided by the expanison of the universe and the lepton asymmetry
is converted to the baryon asymmetry through the weak Sphaleron process.
} in generating the observed baryon asymmetry in the universe by providing an
out-of-equilibrium environment, one of the three Sakharov conditions~\cite{Sakharov:1967dj}, in the framework 
of electroweak baryogenesis (see~\cite{Morrissey:2012db} for a recent review). Augmented with the rapid baryon number 
violating Sphaleron process outside the electroweak bubbles and the CP-violating particle scatterings on the bubble 
walls, a net baryon number can be produced inside the bubbles. Aside from the particle interactions, 
which are used in EWBG calculations, the cosmological context that characterizes the dynamics of the 
EWPT can be calculated from the finite temperature effective potential. 
The standard procedure of 
calculating it includes adding the tree level effective potential, the Coleman-Weinberg 
term~\cite{Coleman:1973jx} and its finite temperature counterpart~\cite{Quiros:1999jp} as well as the 
daisy resummation~\cite{Parwani:1991gq,Gross:1980br}. Since the EWPT in this model is mainly driven
by the cubic terms in the potential and out of concern of a gauge parameter 
dependence~\cite{Patel:2011th} of the effective
potential calculated in the above standard procedure, we take here the high temperature expansion 
approximation, which is gauge invariant, in line with previous 
analyses of this model~\cite{Profumo:2007wc,Profumo:2014opa,Kotwal:2016tex,Huang:2017jws,Alves:2018oct}. This  
effective potential is then given by~\footnote{
We also note that we have neglected a tadpole term proportional to $T^2 s$, which originates
from the $a_1$ and $b_3$ terms in the potential in Eq.~\ref{eq:v}, since it comes with a
factor $v_s/v_{\text{EW}}$ and is suppressed for most of the parameter space giving detectable GWs, 
to be presented in later sections. Indeed its effect has been found to be numerically negligible from 
previous studies~\cite{Profumo:2007wc,Profumo:2014opa}.
}
\begin{eqnarray}
 && V(h,s,T) = - \frac{1}{2} [\mu^2 - \Pi_h(T)] h^2 
  - \frac{1}{2} [-b_2 - \Pi_s(T)] s^2 \nonumber \\
  &&\hspace{0.7cm} + \frac{1}{4} \lambda h^4 + \frac{1}{4} a_1 h^2 s + \frac{1}{4} a_2 h^2 s^2 +
  \frac{b_3}{3} s^3 + \frac{b_4}{4} s^4, \quad \quad
  \label{eq:vt}
\end{eqnarray}
where $\Pi_h(T)$ and $\Pi_s$ are the thermal masses of the fields,
\begin{eqnarray}
  &&  \Pi_h(T) = \left( \frac{2 m_W^2 + m_Z^2 + 2 m_t^2}{4 v^2} + \frac{\lambda}{2} + \frac{a_2}{24} \right) T^2, \nonumber \\
  &&  \Pi_s(T) = \left( \frac{a_2}{6} + \frac{b_4}{4} \right) T^2,
\end{eqnarray}
where the gauge and Yukawa couplings have been written in terms of the physical masses of $W$, $Z$ and 
the $t$-quark. With this effective potential, the thermal history of the EW symmetry breaking can be 
analyzed. It depends mainly on the following key parameters:
\begin{equation}
\centering
T_c, \quad T_n, \quad \alpha, \quad \beta, \quad v_w .
\end{equation}

Here $T_c$ is the critical temperature at which the metastable vacuum and the stable one are degenerate.
Below $T_c$, the phase at the origin in the field space becomes metastable and the new phase becomes 
energetically preferable. The rate at which the tunneling happens is given by~\cite{Turner:1992tz}
\begin{eqnarray}
  \Gamma \sim \mathcal{A}(T) e^{-S_3/T} ,
\end{eqnarray} 
where $S_3$ is the 3-dimensional Euclidean action of the critical bubble, which minimizes the action 
\begin{eqnarray}
S_3(\vec{\phi},T) = 4\pi \int r^2 dr \left[ \frac{1}{2} \left(\frac{d \vec{\phi}(r) }{d r}\right)^2 + V(\vec{\phi},T) \right], 
\end{eqnarray}
and satisfies the bounce boundary conditions
\begin{eqnarray}
  \frac{d \vec{\phi}(r)}{d r}\Big\vert_{r=0} = 0, \quad \quad
  \vec{\phi}(r=\infty) =  \vec{\phi}_{\text{out}} . 
\end{eqnarray}
Here $\vec{\phi}_{\text{out}}$ denotes the two components vev of the fields outside the bubble, which is not necessarily the 
origin for two-step EWPT.  The prefactor $\mathcal{A}(T) \propto T^4$ on dimensional grounds. Its 
precise determination needs integrating out fluctuations around the above static bounce 
solution (see e.g.,~\cite{Dunne:2005rt,Andreassen:2016cvx} for detailed calculations 
or~\cite{Weinberg:1996kr} for a pedagogical introduction).
For the EWPT to complete, a sufficiently large bubble nucleation rate is required to 
overcome the expansion rate. This is quantified as the condition that the probability for a 
single bubble to be nucleated within one horizon volume is ${\cal O}(1)$ at a certain temperature~\cite{Chao:2017vrq}:
\begin{eqnarray}
\int_0^{t_n}  {\Gamma V_{H} (t)}  dt= \int_{T_n}^{\infty} {d T \over T }  \left( 2 \zeta M_{\rm Pl}  \over T  \right)^4 e^{{-S_3/T}} ={\cal O} (1) , \ 
\label{eq:nucleation}
\end{eqnarray}
where $V_H (t)$ is the Horizon volume, $M_{\rm Pl}$ is the Planck mass and $\zeta \sim 3 \times 10^{-2}$. 
From this equation, it follows that $S_3(T)/T \approx 140$~\cite{Apreda:2001us} and the temperature thus solved is defined as 
the nucleation temperature $T_n$.
%%%%%%%%%%%%%%%%%%%%%%%%%%%%%%%%%%%%%%%%%%%%%%%%%%%%%%%%%%%%%%%%%%%%%%%%%%%%%%%%%%%%%%%%%%%%%%%
%%%%%%%%%%%%%%%%%%%%%%%%%%%%%%%%%%%%%%%%%%%%%%%%%%%%%%%%%%%%%%%%%%%%%%%%%%%%%%%%%%%%%%%%%%%%%%%
Expanding the rate at $T_n$, one can define the duration of the EWPT in terms of the inverse of the third parameter $\beta$~\cite{Apreda:2001us}:
\begin{eqnarray}
\beta \equiv H_nT_n \left.{d (S_3/T) \over d T} \right |_{T_n} \; , 
\end{eqnarray}
where $H_n$ is the Hubble rate at $T_n$.

Next, $\alpha$ is the vacuum energy released from the EWPT normalized by the total radiation 
energy density ($\equiv\rho_R$) at $T_n$~\cite{Kamionkowski:1993fg}:
\begin{eqnarray}
  \alpha = \frac{\Delta \rho}{\rho_R} = \frac{1}{\rho_R}\left[-V(\vec{\phi}_b, T) + T \frac{\partial V(\vec{\phi}_b, T)}{\partial T}\right] \Bigg|_{T=T_n} , \
\label{eq:alpha}
\end{eqnarray}
where $\rho_R = g_{\ast}\pi^2 T_n^4/30$ with $g_{\ast} \approx 100$ and
$\vec{\phi}_b$ denotes the two components vev of the broken phase. In this expression, the first term is the free energy from the effective potential 
and the second term denotes the entropy production. Finally, $v_w$ is the bubble wall velocity.

Given that a first order EWPT can proceed and complete, the baryon asymmetry is generated outside the bubbles and then captured by 
the expanding bubble walls. When the EWPT finishes, the universe would be in the EW broken phase with non-zero baryon asymmetry.
To ensure that these baryons would not be washed out, the Sphaleron rate needs to be sufficiently quenched inside the bubbles. This condition
is known as the strongly first order EWPT (SFOEWPT) criterion~\cite{Cline:2006ts,Morrissey:2012db}:
\begin{equation}
  \frac{v_H(T)}{T} \Big|_{T=T_n} \gtrsim 1 .
\end{equation}
The conventional choice of the temperature at which the above condition is evaluated is $T_c$, 
but a more precise timing is the nucleation temperature $T_n$(\blue{see e.g., Ref.~\cite{Grojean:2004xa,Grojean:2006bp,Chala:2018ari}}), which we use here. 
Since generally $T_n < T_c$ and $v_h(T_n) > v_h(T_c)$, it might seem at first glance that the above condition is weaker 
when implemented at $T_n$ than at $T_c$. However the implicit assumption associated with the former requires the capability of the 
EWPT to successfully nucleate, i.e., the condition Eq.~\ref{eq:nucleation} should be satisfied in the 
first place, which is typically a more stringent requirement of the potential.

The presence of two scalar fields gives a richer pattern of EWPT and
makes it possible to complete the EWPT with more than one step~\cite{Patel:2012pi,Ramsey-Musolf:2017tgh,Chao:2017vrq}. 
One can immediately imagine mainly the 
following EWPT types:
\begin{itemize}
\item[(A):] $(0,0) \rightarrow (v_H \neq 0 ,v_S\neq 0)$
\item[(B):] $(0,0) \rightarrow (v_H=0, v_S \neq 0) \rightarrow (v_H\neq 0, v_S \neq 0)$
\item[(C):] $(0,0) \rightarrow (v_H\neq 0, v_S=0) \rightarrow (v_H \neq 0, v_S \neq 0)$
\end{itemize}
where the last vacuum configuration $(v_H \neq 0 ,v_S\neq 0)$ in each case would eventually evolve to the EW vacuum 
at $T=0$~\footnote{More exotic patterns might appear but should be of negligible parameter space. 
For an example, see Ref.~\cite{Angelescu:2018dkk}.}.
Here pattern (A) is a one step EWPT from the origin in field space to the EW symmetry breaking vacuum directly,
due mainly to the negative cubic term in the effective potential. This one step phase transition results in 
a typical GW spectrum as shown in the left panel of Fig.~\ref{fig:gw}.
Quite differently, patterns (B) and (C) are 
two-step EWPT, which differ only in how the vacuum transits for these two steps. For example, in case (B), the universe
first goes to a vacuum which has non-zero vev for the singlet field and then transits to the would-be EW vacuum 
at high temperature. Case (C) is different in that it breaks the EW vacuum first and then further goes to the would-be
vacuum in a subsequent step of phase transition. For each transit of the vacuum, it can be either 
first or second order, depending on whether there is a barrier separating the two vacua. We note that for case (C), baryon production generally
needs to occur in the first step, otherwise, the exponentially reduced Sphaleron rate would greatly suppress the 
baryon number violating process in the second step as the EW symmetry is already broken outside the bubbles. 
Therefore the SFOEWPT criterion is imposed in the first step for this case. 

We note that with the aid of the analytical methods presented in Ref.~\cite{Espinosa:2011ax,Chao:2017vrq}, 
it is possible to locate the region of the parameter space that gives exactly one specific type of EWPT by imposing various 
conditions on the input parameters. However, our task here is to reveal the overall behavior of the parameter 
space concerning EWPT and GW. 
Therefore we adopt here a scan-based analysis which covers the entire parameter space and 
for each scanned parameter space point, we determine its pattern of EWPT and calculate GW properties. 
This way, we can determine the
most probable pattern of EWPT admitted by this model. 

\subsection{\label{sec:hydro}Hydrodynamics}

Successful EWBG usually requires a subsonic $v_w$ to give sufficient time for chiral asymmetry propagation ahead
of the wall and for conversion to baryon asymmetry through the Sphaleron process. 
On the other hand, a larger $v_w$ generally leads to more energy being released to the kinetic energy of 
the plasma and therefore a stronger GW production. 
Therefore a tension may arise between  successful EWBG and a loud GW signal production.
This problem can potentially be solved when the hydrodynamic properties of the fluid are taken into account~\cite{No:2011fi}. This is because the
expanding wall stirs the fluid surrounding the bubble wall and a non-zero velocity profile exists for the plasma ahead of the
wall (see Ref.~\cite{Espinosa:2010hh} for a recent combined analysis). In the bubble wall frame, this means the plasma outside the bubble will head towards the bubble wall with a velocity ($\equiv v_+$) that can be different from $v_w$. Therefore it is $v_+$ rather than 
$v_w$ that should be used in EWBG calculations. While the above argument still needs to be scrutinized
taking into account the particle transport behavior around the bubble wall in the process of EWBG, 
we assume tentatively that this is true in this work.

%%%%%%%%%%%%%%%%%%%%%%%%%%%%%%%%%%%%%%%%%%%%%%%%%%%%%%%%%%%%%%%%%%%%%%%%%%%%%%%%%%%%%%%%%%%%%%%%%%%%%%%%%%%%%%%%%
%%%%%%%%%%%%%%%%%%%%%%%%%%%%%%%%%%%%%%%%%%%%%%%%%%%%%%%%%%%%%%%%%%%%%%%%%%%%%%%%%%%%%%%%%%%%%%%%%%%%%%%%%%%%%%%%%
\begin{figure}[t]
\centering
\includegraphics[width=\textwidth]{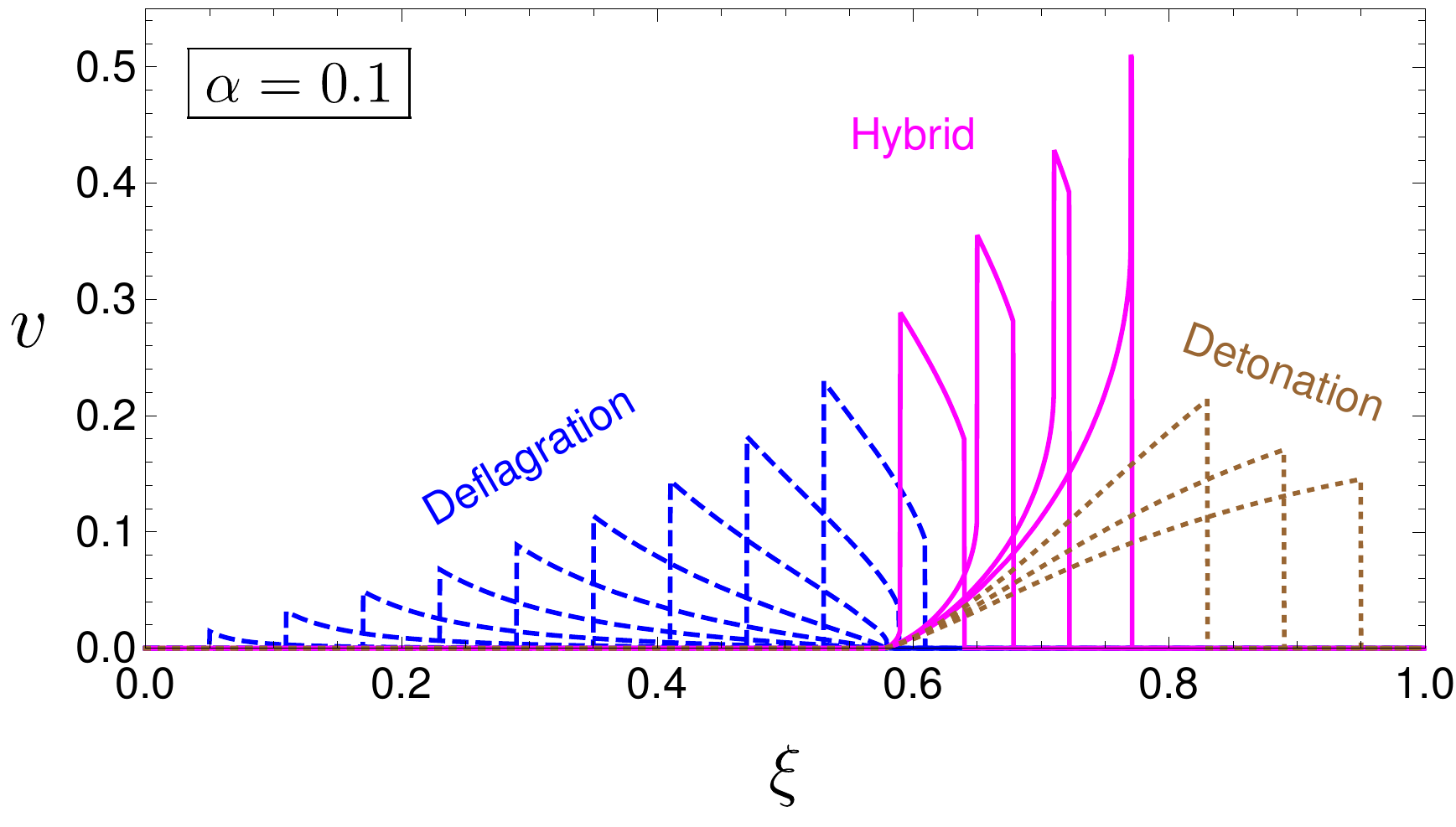}
\caption{\label{fig:modes}
A set of fluid velocity profiles obtained when $v_w$ is increased from small to large values(from left to right), for $\alpha=0.1$.
Three modes of profiles are obtained, deflagration (blue dashed), supersonic deflagration (aka hybrid, magenta solid) and
detonation (brown dotted).
}
\end{figure}
%%%%%%%%%%%%%%%%%%%%%%%%%%%%%%%%%%%%%%%%%%%%%%%%%%%%%%%%%%%%%%%%%%%%%%%%%%%%%%%%%%%%%%%%%%%%%%%%%%%%%%%%%%%%%%%%%
%%%%%%%%%%%%%%%%%%%%%%%%%%%%%%%%%%%%%%%%%%%%%%%%%%%%%%%%%%%%%%%%%%%%%%%%%%%%%%%%%%%%%%%%%%%%%%%%%%%%%%%%%%%%%%%%%

This hydrodynamic treatment hinges on solving the fluid velocity profile $v(r,t)$ around the bubble wall given inputs of $(\alpha, v_w)$,
where $r$ is the distance from the bubble center and $t$ is counted from the onset of the EWPT.
Due to the properties of the problem here, $v$ is a function solely of $r/t \equiv \xi$.
The differential equation governing the velocity profile is derived from the conservation of the energy momentum tensor describing
the fluid and scalar field~\cite{Espinosa:2010hh}:
\begin{equation}
  2 \frac{v}{\xi} = \frac{1-v \xi}{1- v^2}  \left[\frac{\mu^2}{c_s^2} -1 \right] \partial_{\xi} v ,
  \label{eq:velocity}
\end{equation}
where $c_s = 1/\sqrt{3}$ is the speed of sound in the plasma and $\mu(\xi, v) =(\xi-v)/(1-\xi v)$ is a Lorentz boost transformation.
Far outside the bubble and deep inside the bubble, the plasma will not be stirred, that is $v\rightarrow 0$  serves as the boundary 
condition. At the phase boundary, the velocity of the plasma inside and outside the bubble wall are denoted as $v_-$ and $v_+$ in 
the bubble wall
frame, both heading towards the bubble center. The same energy momentum conservation, when applied across the bubble wall, gives a continuity equation connecting $v_-$ with $v_+$. Therefore the whole fluid velocity profile can be solved from the center of the bubble
to far outside the bubble where the plasma is unstirred. 

The solutions of the fluid profiles can be classified into three modes depending on the value of $v_w$. A set of profiles $v(\xi)$ are shown 
in Fig.~\ref{fig:modes} for $\alpha = 0.1$. For $v_w < c_s$, a deflagration mode is obtained, in which case, the plasma ahead of the bubble wall flows outward while it remains 
static inside the bubble, corresponding to the profiles with blue-dashed lines. It can also be seen from this figure that as $v_w$ increases
in this mode, a discontinuity in $v(\xi)$ appears outside the bubble and $v(\xi)$ jumps to zero. 
This is the location of the shock front, and beyond this point the solution of Eq.~\ref{eq:velocity} is 
invalid and a shock front develops such that $v(\xi)$ goes to zero consistently.
When  $v_w$ surpasses $c_s$ but is less than a certain threshold $\xi_J(\alpha)$, a supersonic deflagration mode~\cite{KurkiSuonio:1995pp} 
appears (magenta solid profiles) where the plasma inside the bubble has a non-zero profile, while still taking the form of deflagration 
outside the bubble. Here $\xi_J(\alpha)$, as a function of $\alpha$, 
corresponds to the Jouguet detonation~\cite{Steinhardt:1981ct}, used in earlier studies. It is also evident that in this mode, as $v_w$ increases, the shock front becomes closer to the bubble wall until 
it coincides
with the bubble wall, where $v_w = \xi_J(\alpha)$ and the fluid enters the third, detonation mode (brown dotted profiles). In this mode,
the plasma outside the bubble has zero velocity and therefore $v_+ = v_w$. 
If a subsonic velocity is required in EWBG, we conclude that 
the deflagration mode will not work for EWBG. On the contrary, $v_+ < v_w$ in the deflagration and supersonic deflagration modes and a solution 
for the tension between EWBG and GW might be achieved. 

Therefore, instead of treating $v_w$ as a free parameter in the GW calculations, we require, given a certain input of $\alpha$, the 
corresponding $v_+$ to have subsonic value, taken to be $0.05$ 
here, a choice usually used in EWBG calculations~\cite{John:2000zq,Cirigliano:2006dg,Chung:2009qs,Chao:2014dpa,Guo:2016ixx}).
The procedure of achieving the above goal is as follows: for each given $\alpha$ we iterate over $v_w$ and solve the whole fluid profile 
until $v_+=0.05$ is reached. The resulting $v_w$ is used in GW calculations~\footnote{
For two-step EWPT, a small $v_+$ is not necessarily required for both steps of EWPT. However since $v_w$ is otherwise an almost free parameter,
we stick to the choice $v_+=0.05$ for both steps. 
}.

With $v(\xi)$ obtained, one can also calculate the bulk kinetic energy normalized by the vacuum energy released during the EWPT~\cite{Espinosa:2010hh}:
\begin{eqnarray}
  \kappa_v = \frac{3}{\Delta \rho\ v_w^3} \int \omega(\xi) \frac{v^2}{1-v^2} \xi^2 d \xi ,
  \label{eq:kv}
\end{eqnarray}
where $\omega(\xi)$ is the enthalpy density, varying as function of $\xi$, and can be solved once $v(\xi)$ is found. The remaining part 
$1-\kappa_v \equiv \kappa_T$ gives the fraction of the vacuum energy going to heat the plasma. 
Therefore a reheating temperature can be defined as 
\begin{eqnarray}
T_{\ast} = T_n (1+\kappa_T \alpha)^{1/4}.
\label{eq:Tast}
\end{eqnarray}
This leads to an increase in entropy density and thus a dilution of the generated baryon asymmetry~\cite{Patel:2012pi}. 
Typically in EWBG calculations, the wall curvature is neglected and the transport equations depend on a 
single coordinate $\bar{z}$ in the bubble wall rest frame, where $\bar{z}>0$ ($<0$) corresponds to 
broken (unbroken) phase. The solved baryon asymmetry density $n_B$ is a constant inside the bubbles(see, e.g.,~\cite{Lee:2004we}):
\begin{eqnarray}
  n_B = \frac{3 \Gamma_\mathrm{ws}}{D_q \lambda_+} \int_0^{-\infty}n_L(\bar{z}) e^{-\lambda_- \bar{z}} d\bar{z}\ \  ,
\end{eqnarray}
where $s(T)=2g_{\ast}\pi^2 T^3/45$ is the entropy density, 
$\Gamma_{\text{ws}} \approx 120 \alpha_w^5 T$ is the weak Sphaleron rate in the EW symmetric 
phase~\cite{Bodeker:1999gx}, $\lambda_{\pm} = (v_+ \pm \sqrt{v_+^2 + 15 \Gamma_{\text{ws}}D_q})/(2 D_q)$
with $D_q$ the diffusion constant for quarks~\cite{Bodeker:1999gx} and 
$n_L$ is the chiral asymmetry of left-handed doublet fields which serves as a source term in 
baryon asymmetry generation. The determination of $n_L$ is a key part in EWBG calculations and is 
decoupled from the analysis of EWPT dynamics here. In above expression, we have replaced $v_w$ by $v_+$,
to take into account  the distinction between these two velocities. If the temperature at which 
$n_B$ is calculated is $T_n$, then
after the bubbles have collided, the temperature of the plasma is given, to a good approximation, by $T_{\ast}$  
rather than $T_n$ or $T_c$, which are conventionally used. The diluted baryon asymmetry is then given by
\begin{eqnarray}
  \frac{n_B}{s} |_{T = T_{\ast}} = \xi_D \frac{n_B}{s}|_{T = T_n},
  \label{eq:dilution}
\end{eqnarray}
where $\xi_D \equiv (1+\kappa_T \alpha)^{-3/4}$ captures the dilution effect of the generated baryon 
asymmetry by reheating of the plasma.
We then need to make sure that $\xi_D$ does not become too small, since otherwise a stronger CP-violation will
be needed, which might be excluded by the stringent limits from electric dipole moment 
searches~\cite{Engel:2013lsa,Chupp:2017rkp}.

\subsection{\label{sec:gw}Stochastic Gravitational Waves}

During the EWPT, bubbles of EW broken phase expand and collide with each other, which destroys the 
spherical symmetry of a single bubble, thus leading to the emission of gravitational waves~\cite{Kamionkowski:1993fg}. 
Due to the nature of this process and according to the central limit theorem, the generated gravitational 
wave amplitude is a random variable which is isotropic, unpolarized and follows a Gaussian 
distribution. This therefore allows the description of gravitational wave amplitude using its two-point correlation
function and is parametrized by the gravitational wave energy density spectrum $\Omega_{\text{GW}}(f)$, as
a function of frequency $f$.
A natural consequence is  that the GWs produced during the EWPT, when redshifted to the present, give a peak
frequency at around the milli-Hertz range~\cite{Grojean:2006bp}, falling right within the band of future space-based 
gravitational wave detectors. 

It is now well known that there are mainly three sources of gravitational
wave production in this process: bubble wall collisions~\cite{Kosowsky:1991ua,Kosowsky:1992rz,Kosowsky:1992vn,Huber:2008hg,Jinno:2016vai,Jinno:2017fby}, sound waves in the plasma~\cite{Hindmarsh:2013xza,Hindmarsh:2015qta} and magneto-hydrodynamic turbulence (MHD)~\cite{Hindmarsh:2013xza,Hindmarsh:2015qta}. The total 
energy density spectrum can be obtained approximately by adding these contributions:
\begin{equation}
  \Omega_{\text{GW}}h^{2} \simeq \Omega_{\text{col}}h^{2}+\Omega_{\text{sw}}h^{2}+\Omega_{\text{turb}}h^{2} .
\end{equation}
Recent studies suggest that the energy deposited in the bubble walls is negligible, despite the 
possibility that the bubble walls can run away in some circumstances~\cite{Bodeker:2009qy}. Therefore while a bubble wall can reach  relativistic speed, its contribution to gravitational waves can generally be neglected~\cite{Bodeker:2017cim}. 
We thus include only the contribution of sound waves and turbulence in the gravitational wave spectrum calculations.

The dominant contribution comes from sound waves. 
By evolving the scalar-field and fluid model on 3-dimensional lattice, 
the gravitational wave energy density spectrum can be extracted, with an analytical fit formula 
available~\cite{Hindmarsh:2015qta}:
\begin{eqnarray}
  &&\Omega_{\textrm{sw}}h^{2}=2.65\times10^{-6}\left( \frac{H_{\ast}}{\beta}\right) \left(\frac{\kappa_{v} \alpha}{1+\alpha} \right)^{2} 
\left( \frac{100}{g_{\ast}}\right)^{1/3} \nonumber \\
&&\hspace{1.4cm} \times v_{w} \left(\frac{f}{f_{\text{sw}}} \right)^{3} \left( \frac{7}{4+3(f/f_{\textrm{sw}})^{2}} \right) ^{7/2} \ .
\label{eq:soundwaves}
\end{eqnarray}
%%%%%%%%%%%%%%%%%%%%%%%%%%%%%%%%%%%%%%%%%%%%%%%%%%%%%%%%%%%%%%%%%%%%%%%%%%%%%%%%%%%%%%%%%%%%%%%%%%%%%%%%%%%%%%%%%
%%%%%%%%%%%%%%%%%%%%%%%%%%%%%%%%%%%%%%%%%%%%%%%%%%%%%%%%%%%%%%%%%%%%%%%%%%%%%%%%%%%%%%%%%%%%%%%%%%%%%%%%%%%%%%%%%
\begin{figure}
  \centering
\includegraphics[width=0.49\textwidth]{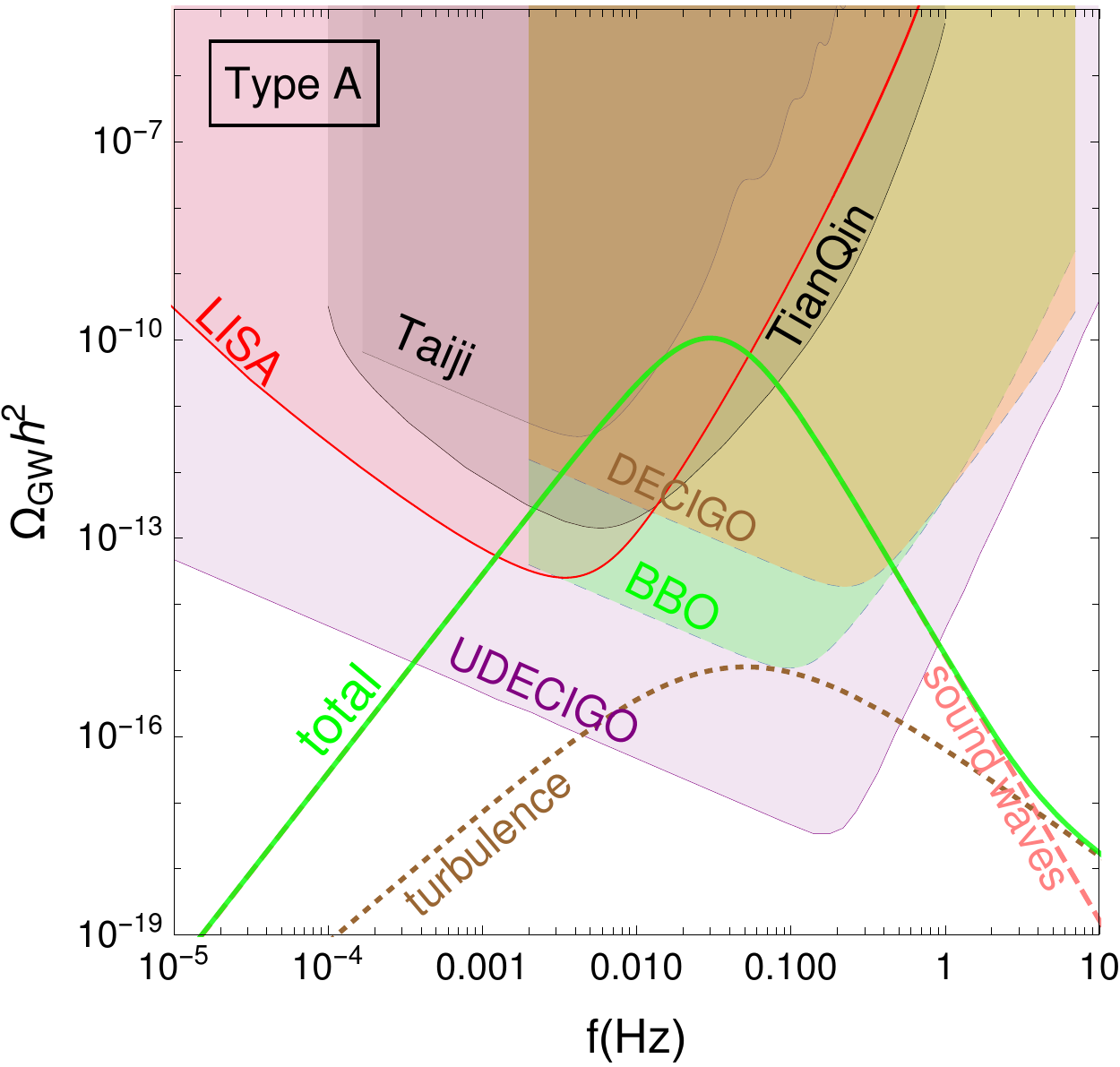}
\ 
\includegraphics[width=0.49\textwidth]{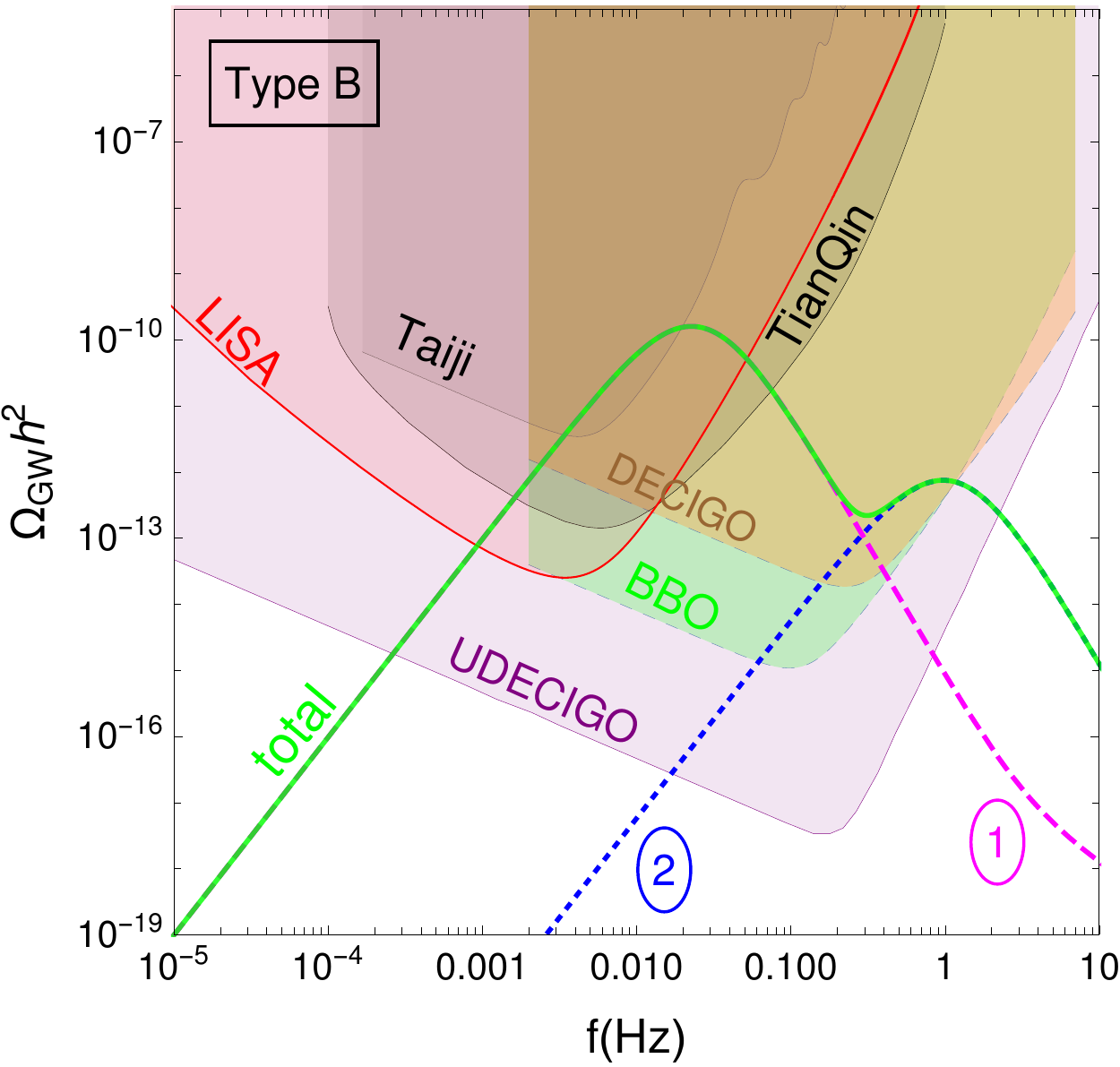}
\caption{
\label{fig:gw}
Examples showing GW energy density spectra from one step (left) and two-step (right) EWPT.
For the left panel, the individual contributions from sound waves and magnetohydrodynamic turbulence
are shown with their sum denoted by the green solid line. For the right panel, the total contributions
from both the first step and second step are shown and with their sum denoted by the green solid line.
}
\end{figure}
%%%%%%%%%%%%%%%%%%%%%%%%%%%%%%%%%%%%%%%%%%%%%%%%%%%%%%%%%%%%%%%%%%%%%%%%%%%%%%%%%%%%%%%%%%%%%%%%%%%%%%%%%%%%%%%%%
%%%%%%%%%%%%%%%%%%%%%%%%%%%%%%%%%%%%%%%%%%%%%%%%%%%%%%%%%%%%%%%%%%%%%%%%%%%%%%%%%%%%%%%%%%%%%%%%%%%%%%%%%%%%%%%%%
%
Here $H_{\ast}$ is the Hubble parameter at $T_{\ast}$ when the phase transition has completed. 
It has a value close to that evaluated at the nucleation temperature $T_n$ for sufficiently short 
EWPT~\cite{Caprini:2015zlo}. We take $T_{\ast}$ to be the reheating temperature, defined earlier 
in Eq.~\ref{eq:Tast}.
Moreover, $f_{\text{sw}}$ is the present peak frequency which is the redshifted value of the peak 
frequency at the time of EWPT ($=2 \beta/(\sqrt{3}v_w)$):
 \begin{equation}
f_{\textrm{sw}}=1.9\times10^{-5}\frac{1}{v_{w}}\left(\frac{\beta}{H_{\ast}} \right) \left( \frac{T_{\ast}}{100\textrm{GeV}} \right) \left( \frac{g_{\ast}}{100}\right)^{1/6} \textrm{Hz} ,
\end{equation}
where $\kappa_{v}$ is defined in Eq.~\ref{eq:kv} and can be calculated as a function of 
($\alpha$, $v_w$) by solving the velocity profiles described in Sec.~\ref{sec:ewpt}~\cite{Espinosa:2010hh}. 
It should be noted that a more recent numerical simulation by the same group~\cite{Hindmarsh:2017gnf,Hindmarsh:2016lnk} 
shows a slightly enhanced $\Omega_{\text{sw}} h^2$ and reduced peak frequency $f_{\text{sw}}$. 
We also note that the results from these simulations are currently limited to regions of small $v_w$ and 
$\alpha$ and therefore their validity for ultra-relativistic $v_w$ and large $\alpha$ (say $\alpha \gtrsim 1$)
remains unknown. In the absence of numerical simulations for these choices of parameters at present, 
we assume that the results shown here apply for these cases and remind the reader to keep the above caveats in mind.

The fully ionized plasma at the time of EWPT can result in the formation of MHD turbulence, which 
gives another source of gravitational waves. The resulting contribution can  also be modelled 
similarly with a fit formula~\cite{Caprini:2009yp,Binetruy:2012ze},
\begin{eqnarray}
  &&\Omega_{\textrm{turb}}h^{2}=3.35\times10^{-4}\left( \frac{H_{\ast}}{\beta}\right) \left(\frac{\kappa_{\text{turb}} 
\alpha}{1+\alpha} \right)^{3/2} \left( \frac{100}{g_{\ast}}\right)^{1/3}  \nonumber \\
  && \hspace{1.8cm}\times v_{w}  \frac{(f/f_{\textrm{turb}})^{3}}{[1+(f/f_{\textrm{turb}})]^{11/3}(1+8\pi f/h_{\ast})} ,
\label{eq:mhd}
\end{eqnarray}
where $f_{\text{turb}}$ is the peak frequency and is given by,
\begin{equation}
f_{\textrm{turb}}=2.7\times10^{-5}\frac{1}{v_{w}}\left(\frac{\beta}{H_{\ast}} \right) \left( \frac{T_{\ast}}{100\textrm{GeV}} \right) \left( \frac{g_{\ast}}{100}\right)^{1/6} \textrm{Hz} .
\end{equation}
Here the factor $\kappa_{\text{turb}}$ describes the fraction of energy transferred to the MHD turbulence 
and is given roughly by $\kappa_{\text{turb}}\approx \epsilon \kappa_{v} $ with 
$\epsilon \approx 5 \sim 10\%$~\cite{Hindmarsh:2015qta}. We take $\epsilon = 0.1$ in this study.

In both Eq.~\ref{eq:soundwaves} and ~\ref{eq:mhd}, 
the value of $v_w$ is found by requiring that $v_+=0.05$ by solving the velocity profiles, as 
discussed in the previous section.
For the two-step EWPT, as discussed in last section, if both steps in case (B) and (C) are first order, 
then there would be two subsequent GW generation at generally different
peak frequencies and amplitudes, corresponding to the example shown in the right panel of Fig.~\ref{fig:gw}.

The detectability of the GWs is quantified by the signal-to-noise ratio (SNR), whose definition is given in 
Ref.~\cite{Caprini:2015zlo}: 
%%%%%%%%%%%%%%%%%%%%%%%%%%%%%%%%%%%%%%%%%%%%%%%%%%%%%%%%%%%%%%%%%%%
\begin{eqnarray}
  \text{SNR} = \sqrt{\delta \times \mathcal{T} \int_{f_{\text{min}}}^{f_{\text{max}}} df 
    \left[
      \frac{h^2 \Omega_{\text{GW}}(f)}{h^2 \Omega_{\text{exp}}(f)} 
  \right]^2} .
\end{eqnarray}
Here $h^2 \Omega_{\text{exp}}(f)$ is the experimental sensitivity and corresponds to the lower boundaries
of the color-shaded regions in Fig.~\ref{fig:gw} for the shown detectors~\footnote{
There are possible astrophysical foregrounds coming from, e.g., the superposition of 
unresolved (i.e., low SNR) gravitational wave signals of the white dwarf binaries in our 
Galaxy~\cite{Klein:2015hvg}. Including these will slightly reduce the SNR calculated here.
}.
$\mathcal{T}$ is the mission duration in years for each experiment, assumed to be $5$ here. 
The factor $\delta$ comes from the number of independent channels for cross-correlated 
detectors, which equals $2$ for BBO as well as UDECIGO and $1$ for the others~\cite{Thrane:2013oya}.
In our numerical analysis, we stick to the most mature LISA detector with the C1 configuration, 
defined in Ref.~\cite{Caprini:2015zlo}.
To qualify for detection, the SNR needs to be larger than a threshold value, which depends on
the details of the detector configuration. For example, for a four-link LISA configuration, 
the suggested value is 50 while for a six-link configuration, this value can be much lower 
($\text{SNR}=10$), since in this case a special noise reduction technique is 
available based on the correlations of outputs from the independent sets of interferometers
of one detector~\cite{Caprini:2015zlo}.

As an example, we scan over the EW vacuum stability regions in the plane
$(v_s/v_{\text{EW}},b_3/v_{\text{EW}})$ of Fig.~\ref{fig:demo} 
and found the regions which can give successful bubble nucleations, satisfy the SFOEWPT criterion and 
generate GWs. These regions are plotted with blue ($\text{SNR}<10$), green ($50>\text{SNR}>10$) and 
red ($\text{SNR}>50$). Here most of the points give type (A) EWPT with only several
points for type (B) or (C), denoted by diamond shapes.

\section{\label{sec:result}Results and Discussions}

In this section, we perform a full scan of the parameter space to address the following questions:
\begin{itemize}
\item[(a)] What kind of EWPT patterns can this model admit and in what proportion of the parameter space for each pattern?
\item[(b)] What is the region of parameter space that can give strong detectable gravitational waves at future 
space-based gravitational wave detectors?
\item[(c)] Do current collider measurements of double Higgs production and searches for a heavy resonance 
decaying to weak boson pairs exclude the points that give strong gravitational 
waves and could future high luminosity LHC (HL-LHC) at $3 \text{ab}^{-1}$ probe the parameter space giving strong gravitational
waves?
\item[(d)] How will a future space-based gravitational wave experiment  complement current and future 
  searches for a heavy scalar resonance?
\end{itemize}

%%%%%%%%%%%%%%%%%%%%%%%%%%%%%%%%%%%%%%%%%%%%%%%%%%%%%%%%%%%%%%%%%%%%%%%%%%%%%%%%%%
We find that it is more efficient to cover the parameter space in the scan using the tadpole basis parameters. 
So we start the random number generating in the tadpole basis. Once a parameter space point is obtained from the sampling, 
it is converted to the non-tadpole basis parameters and for all subsequent phenomenological checks and calculations. The ranges of the 
tadpole basis parameters where we do the random sampling are the following:
\begin{eqnarray}
&& \hspace{0cm} b_4 \in [0.001,5], \quad \quad b_3/v_{\text{EW}} \in [-10,10], \quad   \nonumber \\
  &&a_2 \in [-2 \sqrt{\lambda b_4}, 25], \quad \theta \in [-0.35,0.35], \nonumber \\
  && m_{h_2}\in [260,1000] ,
\end{eqnarray}
where the lower range of $a_2$ is determined by the requirement that the potential is bounded from below. The scan
takes into account  the previously discussed theoretical and phenomenological requirements. Points which pass
these selection criteria are fed into a modified version of \texttt{CosmoTransitions}~\cite{Wainwright:2011kj} 
for calculating the thermal history and the parameters
relevant for EWPT~\footnote{
Other packages include \texttt{Bubbleprofiler}~\cite{Athron:2019nbd} and \texttt{AnyBubble}~\cite{Masoumi:2016wot}.
It should also be noted that there generically exists a difficulty for solving bounce solutions in very thin-walled cases, 
the discussion of which can be found in above paper of \texttt{Bubbleprofiler} and also in Ref.~\cite{Piscopo:2019txs} where 
the neural networks is introduced to solve the bounce solutions. We have verified that the majority of the points we used
are not very thin-walled.
}. Those which can give a successful EWPT by meeting the bubble nucleation criteria are 
further scrutinized for the EWPT type and SFOEWPT conditions. The final remaining
points are used to calculate the gravitational wave spectra, the SNR and collider observables. 

%%%%%%%%%%%%%%%%%%%%%%%%%%%%%%%%%%%%%%%%%%%%%%%%%%%%%%%%%%%%%%%%%%%%%%%%%%%%%%%%%%%%%%%%%%%%%%%%%%%%%%%%%%%%%%%%%
%%%%%%%%%%%%%%%%%%%%%%%%%%%%%%%%%%%%%%%%%%%%%%%%%%%%%%%%%%%%%%%%%%%%%%%%%%%%%%%%%%%%%%%%%%%%%%%%%%%%%%%%%%%%%%%%%
\begin{figure*}[t]
\includegraphics[width=0.33\textwidth]{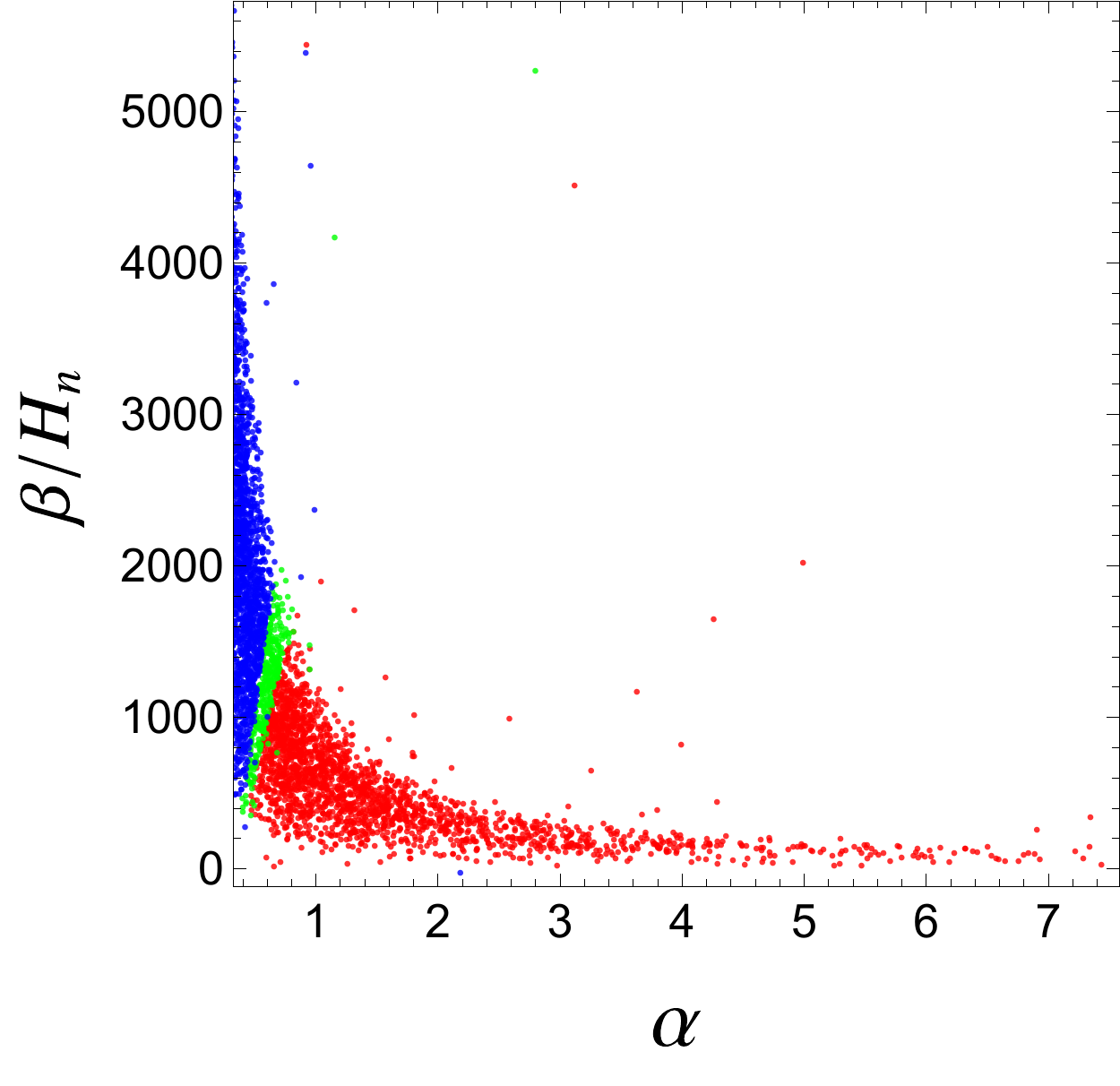}
\includegraphics[width=0.328\textwidth]{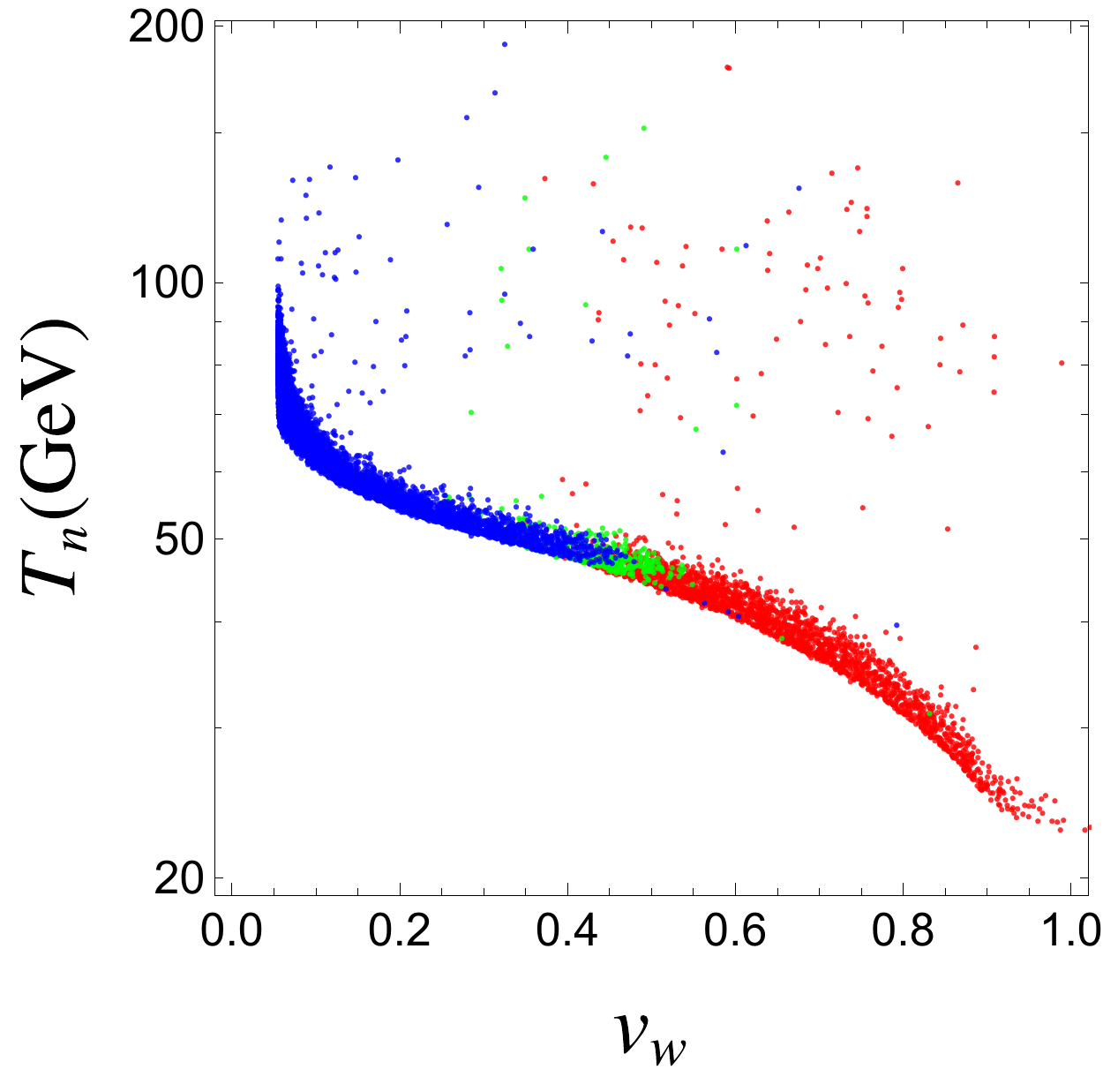} 
\includegraphics[width=0.325\textwidth]{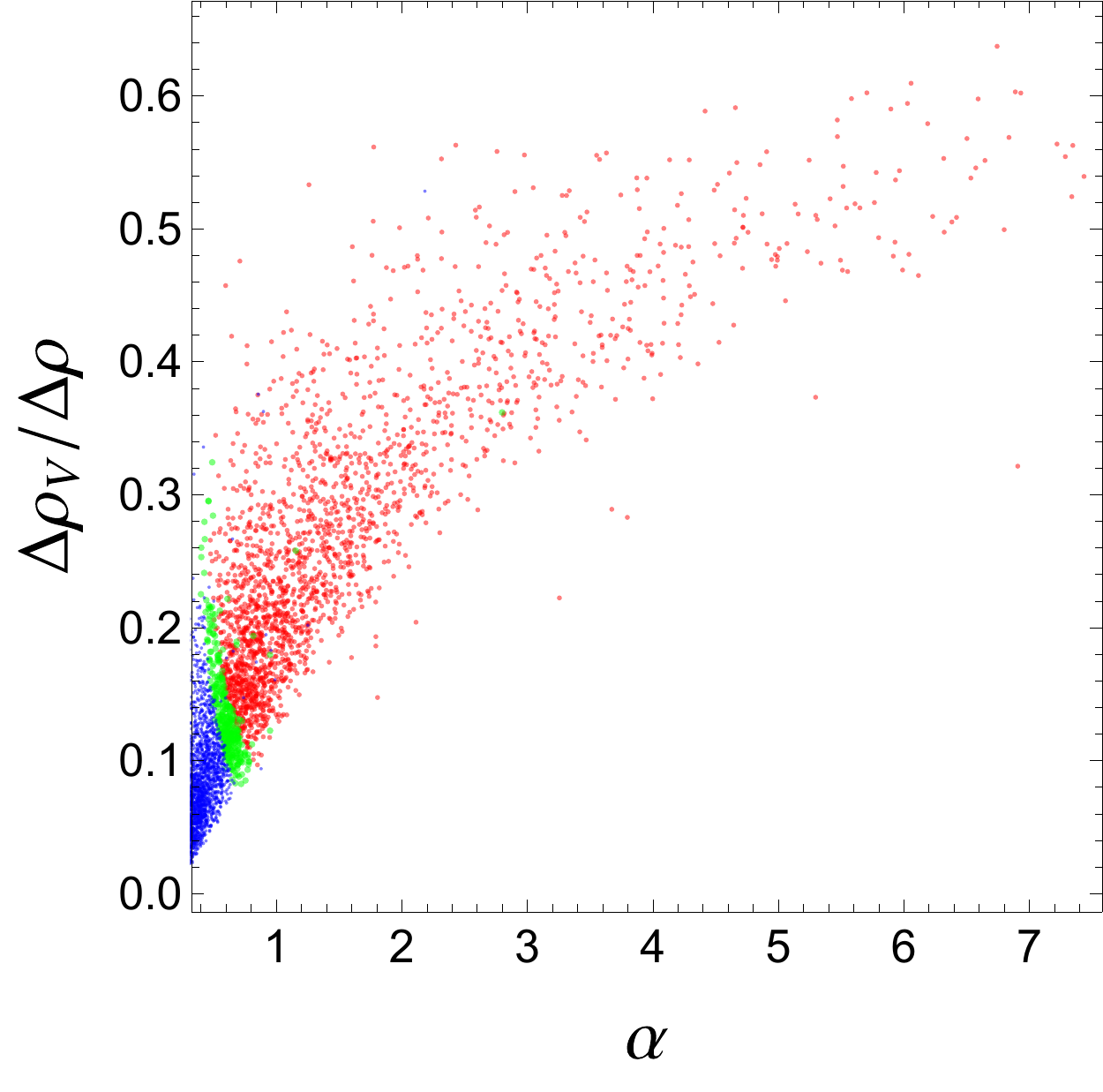}
\caption{\label{fig:scansummary}
The physical parameters characterizing the dynamics of the EWPT: in the plane of $(\alpha, \beta/H_n)$ (left), 
  $(v_w,T_n)$ (middle) and ($\alpha, \Delta \rho_V/\Delta \rho$)  (right). 
In all these plots, the colors denote $\text{SNR}>50$ (red), $50>\text{SNR}>10$ and $\text{SNR}<10$ (blue). Points depicted here pass all phenomenological constraints and give successful bubble nucleations.
}
\end{figure*}
%%%%%%%%%%%%%%%%%%%%%%%%%%%%%%%%%%%%%%%%%%%%%%%%%%%%%%%%%%%%%%%%%%%%%%%%%%%%%%%%%%%%%%%%%%%%%%%%%%%%%%%%%%%%%%%%%
%%%%%%%%%%%%%%%%%%%%%%%%%%%%%%%%%%%%%%%%%%%%%%%%%%%%%%%%%%%%%%%%%%%%%%%%%%%%%%%%%%%%%%%%%%%%%%%%%%%%%%%%%%%%%%%%%
\subsection{EWPT and GW}

We first give  the answer to question (a): \textit{what kind of EWPT patterns can this model admit and in what proportion of the parameter space for each pattern ?}

We find, of the xSM parameter space where a successful EWPT can be obtained, about $99\%$ gives  type (A) EWPT and the remaining
slightly less than $1\%$ can give  type (B) EWPT. We do not observe type (C) EWPT. For type (A), $22\%$ ($19\%$) gives SNR larger
than $10$ ($50$). So there is a sufficiently large parameter space which can give detectable GW production.

The strength of the stochastic GW background is mainly governed by the two parameters $\alpha$ and $\beta/H_n$, where a
larger $\alpha$ and a smaller $\beta/H_n$ gives stronger GW SNR, as shown in the left panel of 
Fig.~\ref{fig:scansummary}, where
the colors denote $\text{SNR}<10$ (blue), $50>\text{SNR}>10$ (green) and $\text{SNR}>50$ (red). We observe that the points which
give detectable GWs lie in the bottom right region of the population.

Physically, $\alpha$ quantifies the amount of energy released during the EWPT and therefore a larger $\alpha$ gives stronger
GW signals. In addition, for fixed $v_w$, a larger $\alpha$ leads to a larger fraction of energy transformed into the plasma kinetic
energy, quantified by $\kappa_v$, and therefore a further gain in GW production. A further enhancement for larger $\alpha$
comes from the fact that since we fixed $v_+=0.05$, increasing $\alpha$ also increases $v_w$.
It should be noted, even without an explicit calculation, that for each fixed value of $\alpha$, the allowed values of $v_w$ are limited to a certain range (see e.g., Fig. 1 in Ref.~\cite{Alves:2018oct}). This comes from two considerations: (1) admitting consistent hydrodynamic solutions of the plasma imposes a lower
limit on $v_w$; (2) $v_w$ larger than $\xi_J(\alpha)$ gives a detonation mode of the velocity profile, in which case $v_w = v_+ > c_s$ and
therefore $v_+$ is too large for EWBG to work.
We further note that for $\alpha \gtrsim 1$ and $v_w \sim 1$, the calculations of the GW spectra may become unreliable for the following reasons: 
($i$) While the study of Ref.~\cite{Bodeker:2017cim} suggests that the energy stored in the scalar field kinetic energy is negligible,
a very large $\alpha$ might lead to a non-negligible contribution from the bubble collisions. Therefore a better understanding
of the energy budget for this region is needed; ($ii$) the numerical simulations are all performed for 
relatively small $\alpha$ as well as $v_w$ and thus the use of these results for large $\alpha$ and $v_w$ may not be applicable;
($iii$) The universe is no longer radiation dominated at the EWPT but rather vacuum energy dominated. This has the consequence that
bubbles might never meet to finish the EWPT and the universe would be trapped in the metastable phase (see Ref.~\cite{Ellis:2018mja} for a recent analysis).
Despite these issues, we find $49\%$ of points with $\text{SNR}>10$ 
have $\alpha <1$ and removing the points with $\alpha > 1$ does not change the main findings of our work.

%%%%%%%%%%%%%%%%%%%%%%%%%%%%%%%%%%%%%%%%%%%%%%%%%%%%%%%%%%%%%%%%%%%%%%%%%%%%%%%%%%%%%%%%%%%%%%%%%%%%%%%%%%%%%%%%%
%%%%%%%%%%%%%%%%%%%%%%%%%%%%%%%%%%%%%%%%%%%%%%%%%%%%%%%%%%%%%%%%%%%%%%%%%%%%%%%%%%%%%%%%%%%%%%%%%%%%%%%%%%%%%%%%%
\begin{figure}[t]
\centering
\includegraphics[width=0.45\textwidth]{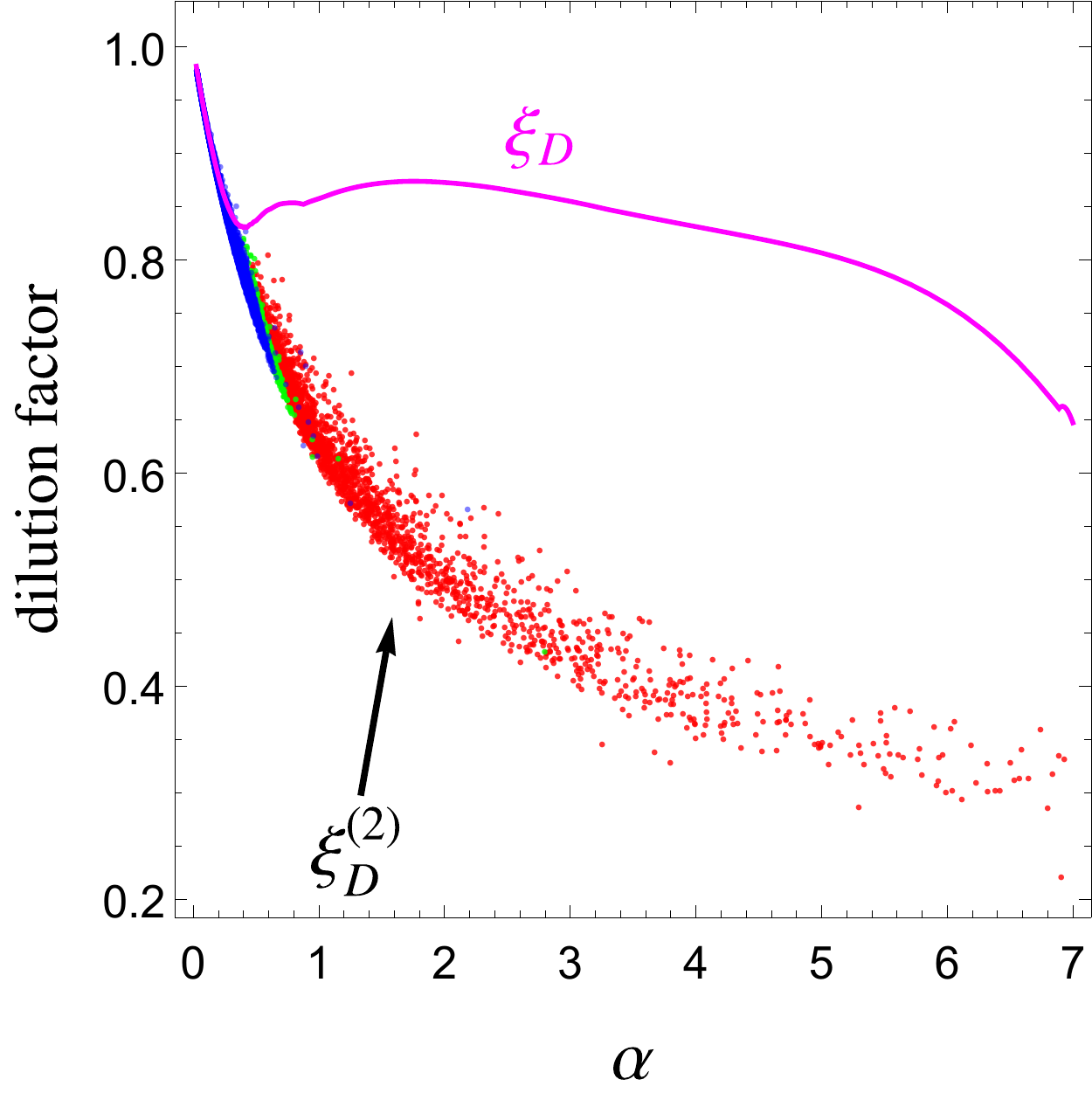}
\includegraphics[width=0.48\textwidth]{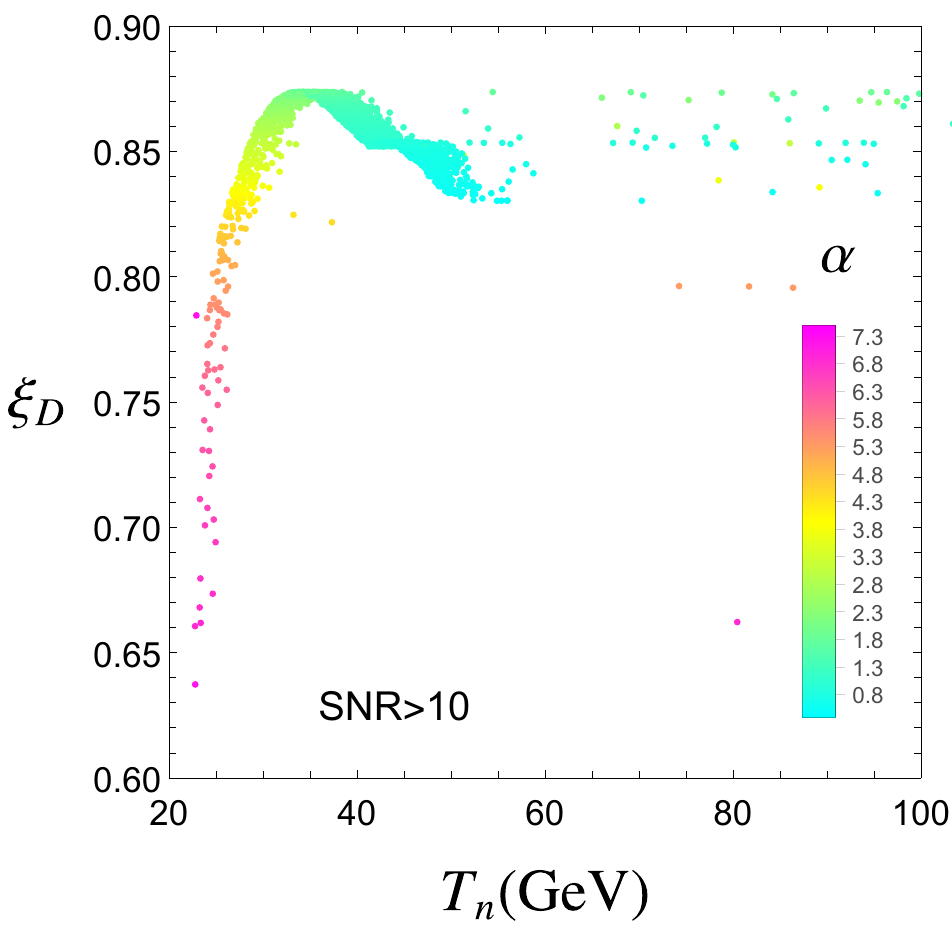}
\caption{
\label{fig:dilution}
Figures showing the dilution effect of the baryon asymmetry.  
The left panel shows two different definitions of the dilution factor and the 
right panel shows the dilution factor $\xi_D$ defined in Eq.~\ref{eq:dilution} versus 
$T_n$.
}
\end{figure}
%%%%%%%%%%%%%%%%%%%%%%%%%%%%%%%%%%%%%%%%%%%%%%%%%%%%%%%%%%%%%%%%%%%%%%%%%%%%%%%%%%%%%%%%%%%%%%%%%%%%%%%%%%%%%%%%%
%%%%%%%%%%%%%%%%%%%%%%%%%%%%%%%%%%%%%%%%%%%%%%%%%%%%%%%%%%%%%%%%%%%%%%%%%%%%%%%%%%%%%%%%%%%%%%%%%%%%%%%%%%%%%%%%%

We now turn to the parameter $\beta/H_n$, which roughly characterizes the inverse time duration of the EWPT. A smaller $\beta/H_n$ or equivalently a longer EWPT generates
stronger GW signals. This is due to the particular feature of the GWs coming from the sound waves in the 
plasma. As was found in the original papers on the importance of sound waves in generating the 
GWs~\cite{Hindmarsh:2013xza,Hindmarsh:2015qta}, one enhancement comes from $1/(\beta/H_n)$ compared with the 
conventional bubble collision contribution. As long as the mean square fluid velocity of the plasma is 
non-negligible, GWs will continue being generated and the energy density of the GW is thus proportional to the 
duration of the EWPT. It should be noted that $\beta/H_n$ also determines the peak frequency of the GW 
spectra.

The bubble wall velocity $v_w$ also plays an important role here and 
the dependence of the SNR on $v_w$ is shown in the middle panel of Fig.~\ref{fig:scansummary}, where the vertical 
axis is chosen to be $T_n$. It is clear that points with larger SNR have larger $v_w$ since, for fixed $v_+$, a larger
$\alpha$ implies a larger $v_w$. It can also be seen from this plot that the SNR increases as $T_n$ decreases. This is
easily understood, since a smaller $T_n$ typically implies a larger amount of supercooling and therefore a larger $\alpha$.
The supercooling can be quantified by the fraction of the first term($\equiv \Delta \rho_V$) of Eq.~\ref{eq:alpha}
in the total released vacuum energy, which we plot in the right panel. 
We can see from this figure that larger SNR indeed implies larger amount of supercooling. However the
amount of supercooling as quantified by $\Delta \rho_V/\Delta \rho$ is less than $0.6$ for most of the parameter space.
The remaining part comes from the second term of the definition of $\alpha$. 

The entropy production, if sizeable, can pose a problem for baryon asymmetry generation, 
as it will effectively dilute the baryon asymmetry $n_B/s$ by increasing $s$. 
In Sec.~\ref{sec:hydro}, we encode this effect in a dilution factor $\xi_D$. Here since $\kappa_T$ is a function of 
$v_w$ and $\alpha$ while $v_w$ is also a function of $\alpha$ when $v_+$ is fixed, we find 
$\xi_D$ is solely
a function of $\alpha$. This functional relation is shown as the magenta line in the left panel of Fig.~\ref{fig:dilution}
and all points from the scan fall on this line. 
The message from this figure  is that most of the points have $\xi_D \gtrsim 0.65$ and those with a smaller $\alpha$ have 
a dilution factor closer to $1$. In particular, the points with $\alpha \lesssim 1$ for which GW can 
be reliably calculated, the dilution effect is rather small as $\xi_D \gtrsim 0.8$. Given the current relatively large 
uncertainties in the EWBG calculations, the dilution effect poses no real problem for the baryon asymmetry generation. 
Note that previous studies~\cite{Patel:2012pi} used a different quantification of the dilution factor, with the definition:
\begin{eqnarray}
  \xi_D^{(2)} = \frac{s}{s + \Delta s} ,
\end{eqnarray}
where $s$ is the entropy density at $T_n$ and $\Delta s$ is calculated from the 
second term in the definition of $\alpha$ in Eq.~\ref{eq:alpha}. To compare with the factor $\xi_D$, what we use here, 
we show values of this factor in the same plot of $\xi_D$ for every point that gives detectable GWs. 
It is evident from this figure that these two factors are roughly the same and both decrease linearly for $\alpha \lesssim 0.4$.
For $\alpha \gtrsim 0.4$, $\xi_D^{(2)}$ gives an overestimation of the dilution effect while $\xi_D$ firstly increases a little
bit before slowly dropping. Since the dilution factor we use here is based on a faithful hydrodynamic analysis, it gives 
a more precise description of the dilution effect. We also show $\xi_D$ calculated for all the points 
versus $T_n$ as a 
scatter plot in the right panel of Fig.~\ref{fig:dilution}, from which we find a larger dilution effect appears for typically smaller $T_n$ and those 
with $\alpha \lesssim 1$ fall in the high $T_n$ region.
 
The two-step EWPT, for which type (B) is the only observed here, constitutes about one percent of all the surviving parameter
space. Of this tiny parameter space, more than half the points give detectable GWs.

%%%%%%%%%%%%%%%%%%%%%%%%%%%%%%%%%%%%%%%%%%%%%%%%%%%%%%%%%%%%%%%%%%%%%%%%%%%%%%%%%%%%%%%%%%%%%%%%%%%%%%%%%%%%%%%%%
%%%%%%%%%%%%%%%%%%%%%%%%%%%%%%%%%%%%%%%%%%%%%%%%%%%%%%%%%%%%%%%%%%%%%%%%%%%%%%%%%%%%%%%%%%%%%%%%%%%%%%%%%%%%%%%%%
\begin{figure}[t]
  \centering
\includegraphics[width=0.31\textwidth]{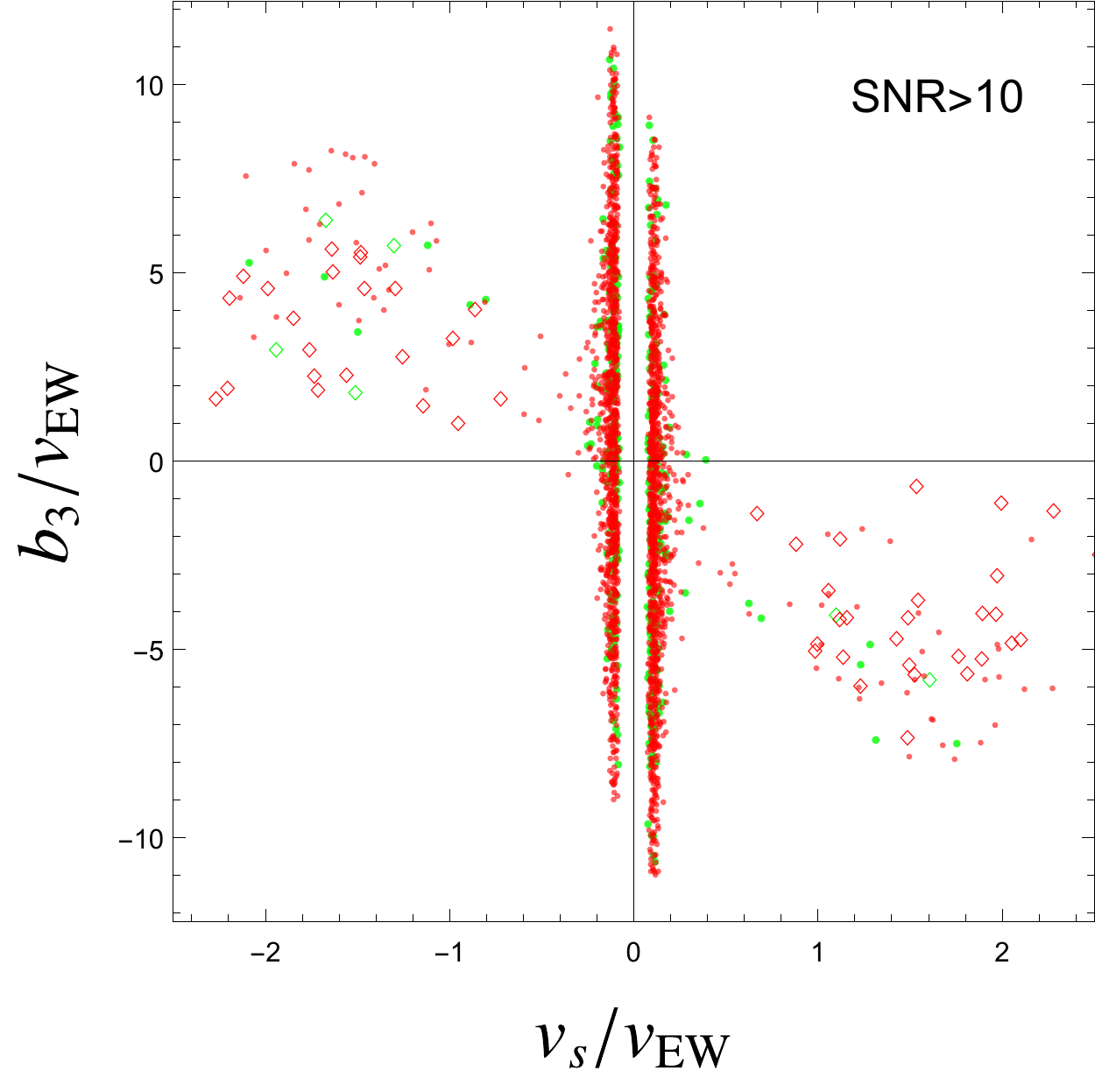}
\includegraphics[width=0.324\textwidth]{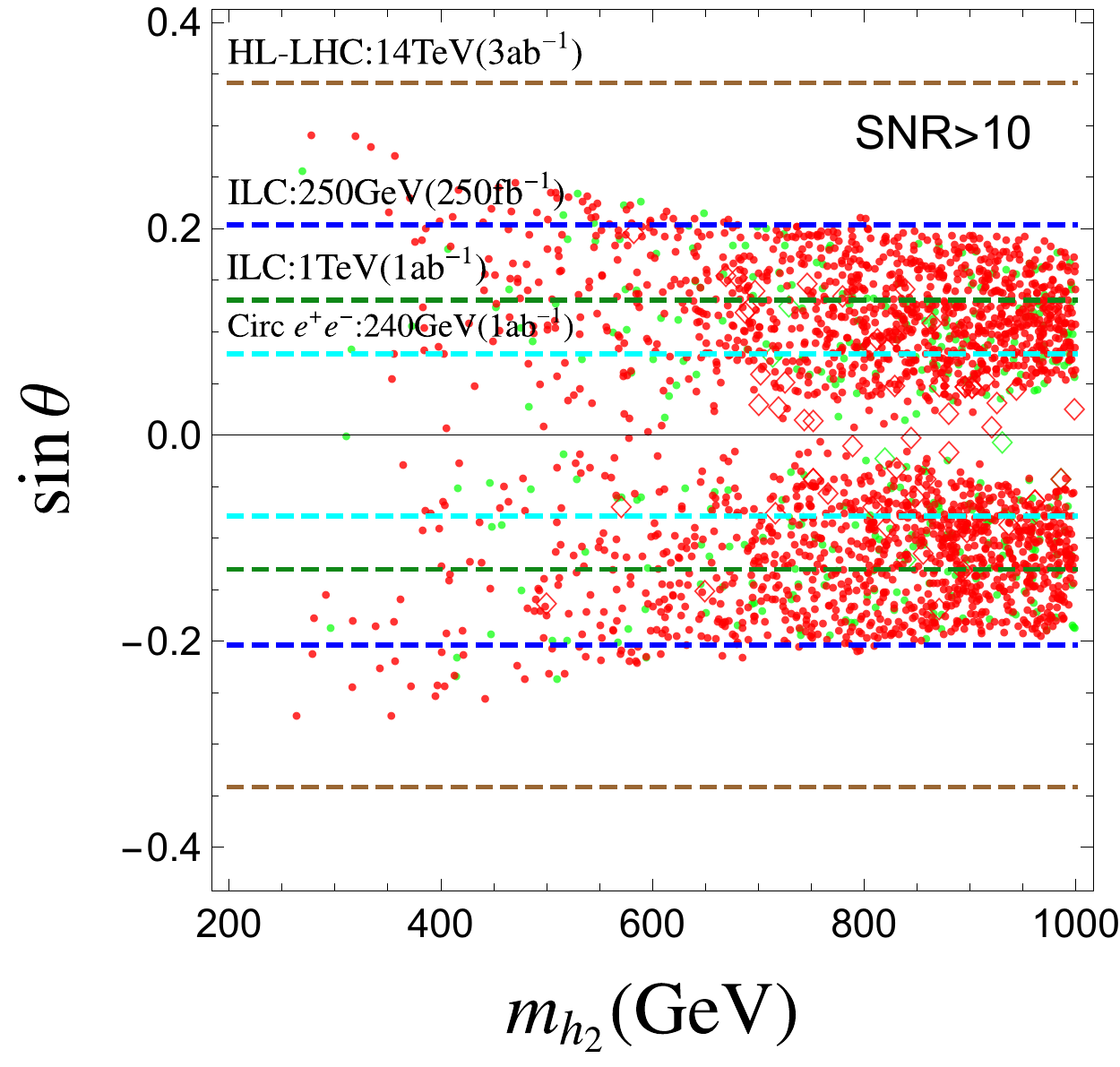}
\includegraphics[width=0.32\textwidth]{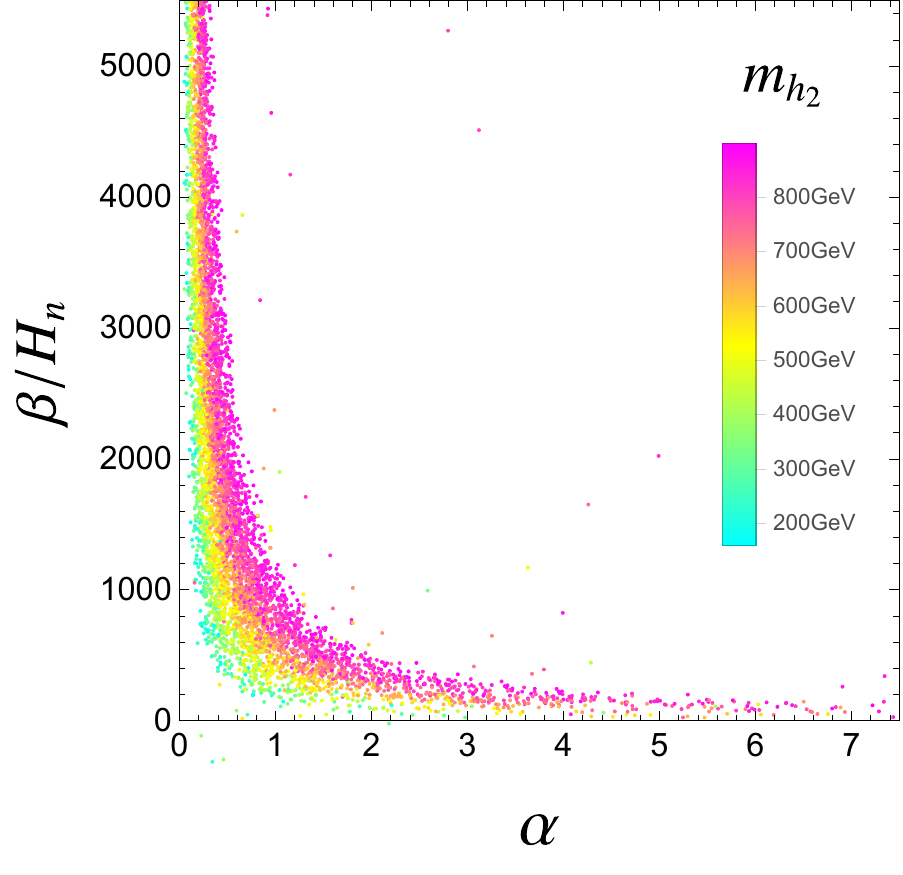}
\caption{\label{fig:parameters}
Points depicted here pass all phenomenological constraints and give successful bubble nucleations, along with detectable GWs at LISA ($\text{SNR}>10$). We show them in the planes of the input parameters: in plane $(b_3,v_s)/v_{\text{EW}}$ (left) and $(\sin\theta, m_{h_2})$ (middle). We distinguish 
those points which give $\text{SNR}>50$ (red) with those of $50>\text{SNR}>10$  (green) in these two plots. The right panel shows 
all the points in the plane $(\alpha,\beta/H_n)$ with the colors denoting the values of $m_{h_2}$, as shown in the legend.
}
\end{figure}
%%%%%%%%%%%%%%%%%%%%%%%%%%%%%%%%%%%%%%%%%%%%%%%%%%%%%%%%%%%%%%%%%%%%%%%%%%%%%%%%%%%%%%%%%%%%%%%%%%%%%%%%%%%%%%%%%
%%%%%%%%%%%%%%%%%%%%%%%%%%%%%%%%%%%%%%%%%%%%%%%%%%%%%%%%%%%%%%%%%%%%%%%%%%%%%%%%%%%%%%%%%%%%%%%%%%%%%%%%%%%%%%%%%
\subsection{Parameter Space Giving Detectable GWs}

With a summary of the points described in previous section, we give in this section the answer to 
question (b), which, we recall, was: \textit{
What is the region of parameter space that can give strong detectable gravitational waves at future 
space-based gravitational wave detectors?}

The results are shown in terms of the three plots in 
Fig.~\ref{fig:parameters}.
As was discussed in the previous section, a large $\alpha$ and small $\beta/H_n$ leads to loud GW signals. 
Even though the relation between $(\alpha, \beta/H_n)$ and the physical input parameters is not 
transparent as many numerical details are involved, it can still be revealed by the 
plots in Fig.~\ref{fig:parameters}.
From the left panel in Fig.~\ref{fig:parameters}, we can see that the majority of the points are concentrated in two 
regions of parameter space where $v_s$ is rather small. In particular, we find $20\, \text{GeV}\lesssim |v_s| \lesssim 50\,\text{GeV}$ 
for most points, with a peak distribution at around $20\,\text{GeV}$.
The appearance of two regions comes from the bounded-from-below
requirement of the potential, similar to Fig.~\ref{fig:demo}. While phenomenological constraints have the effect of shrinking both the regions,
the appearance of points far outside the two regions indeed shows that the main cause of the narrow regions comes from the 
requirements of EWPT and GWs. Therefore it is fair  to say that the region that gives detectable GWs from a type (A) EWPT mainly
comes from the parameter space with smaller $v_s$. On the other hand, the regions which provide type 
(B) EWPT are dramatically different from these regions, since most of the diamonds lie beyond the two 
narrow regions, as can be seen from the figure.

The middle figure shows these regions in the $(m_{h_2}, \sin\theta)$ plane. It is clear that the points are concentrated around
the region with larger $m_{h_2}$. For smaller $m_{h_2}$, the density of points becomes much smaller. To have a better understanding
of the role of $m_{h_2}$ in GW production, we show in the right panel its role in determining $(\alpha, \beta/H_n)$, denoted by
the colors. In this figure, the points are separated into different bands characterized by the value of $m_{h_2}$. For fixed
$\beta/H_n$, a larger $m_{h_2}$ gives a larger $\alpha$, thus larger SNR. This explains the concentration of the points in the
$m_{h_2}$ direction in the middle figure.
In the $\sin \theta$ direction,
the value of $\theta$ is more constrained for larger $m_{h_2}$. The outer boundary comes mainly from the 
$W$-mass constraint. The requirements from EWPT and larger GW signals also show their effects in this 
plot. For example, very small values of $\theta$ give rarer points. 
We also overlaid on this plot the various sensitivity 
projections from colliders in probing the value of $\theta$, which includes HL-LHC, ILC with two configurations (ILC-1: $250\text{GeV}$,
$250\text{fb}^{-1}$, ILC-3: $1\text{TeV}$,$1\text{ab}^{-1}$) and future circular $e^+e^-$ colliders ($240\text{GeV},1\text{ab}^{-1}$), 
all taken from Ref.~\cite{Profumo:2014opa}.
We see that HL-LHC can barely probe any points; ILC-1 can probe a fraction of the small $m_{h_2}$ points as well as a few large $m_{h_2}$ points; 
ILC-3 can probe about a half of both light and heavy $h_2$ points; the future circular colliders can probe even more of the parameter space.
We also can see that most of the points coming from the two-step EWPT lie at the very small $\theta$ region, even though a few do have larger $\theta$.
Therefore GW detections serve as a complementary probe of this region. We also note that for very small values of $\theta$ and $m_{h_2}$,
the search for long lived particles can be used to probe this region (eg., the MATHUSLA detector)~\cite{Curtin:2018mvb}.

\subsection{Correlation with Double Higgs Production Searches}

Exploring possible deviations from the expected SM value of the cubic Higgs coupling through di-Higgs production is an important target of the HL-LHC. New physics scenarios, especially those designed for providing a SFOEWPT for baryon asymmetry generation, typically
modify this coupling. Therefore di-Higgs production is correlated with  EWPT and thus GW production. Future GW and 
collider experiments can then operate in a way that complement each other in exploring new physics scenarios. 
With the parameter space giving detectable GW identified in the previous section, we can find the correlation by calculating
the corresponding di-Higgs cross sections and compare it with present di-Higgs measurements and with future projections.

%%%%%%%%%%%%%%%%%%%%%%%%%%%%%%%%%%%%%%%%%%%%%%%%%%%%%%%%%%%%%%%%%%%%%%%%%%%%%%%%%%%%%%%%%%%%%%%%%%%%%%%%%%%%%%%%%
\begin{figure}[h]
\centering
  \includegraphics[width=0.3\columnwidth]{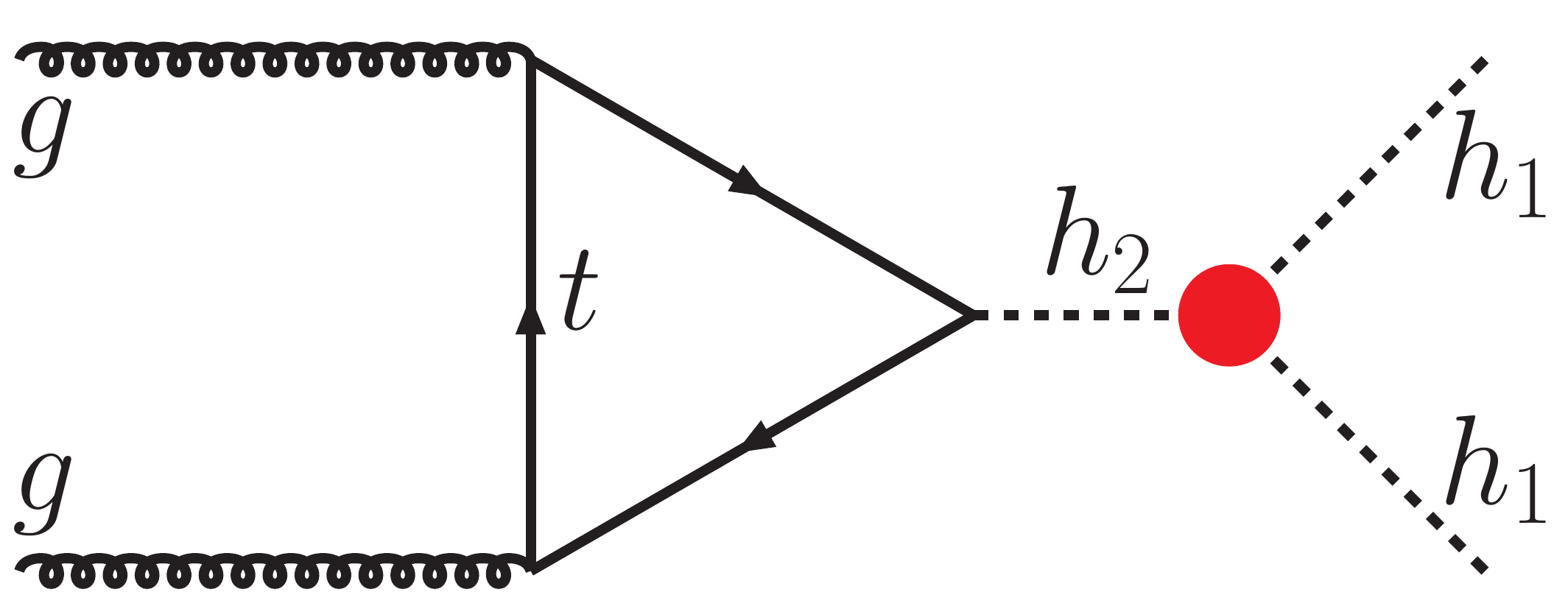}
  \quad
  \includegraphics[width=0.3\columnwidth]{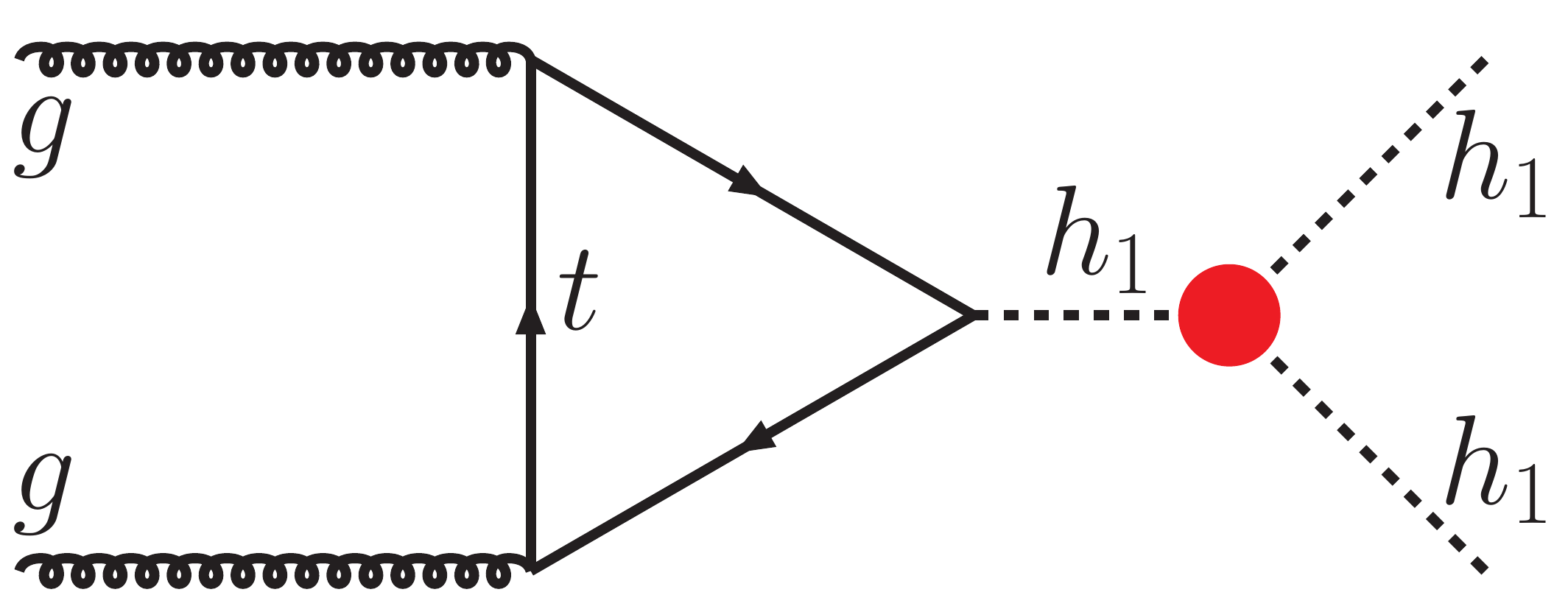}
  \quad
  \includegraphics[width=0.3\columnwidth]{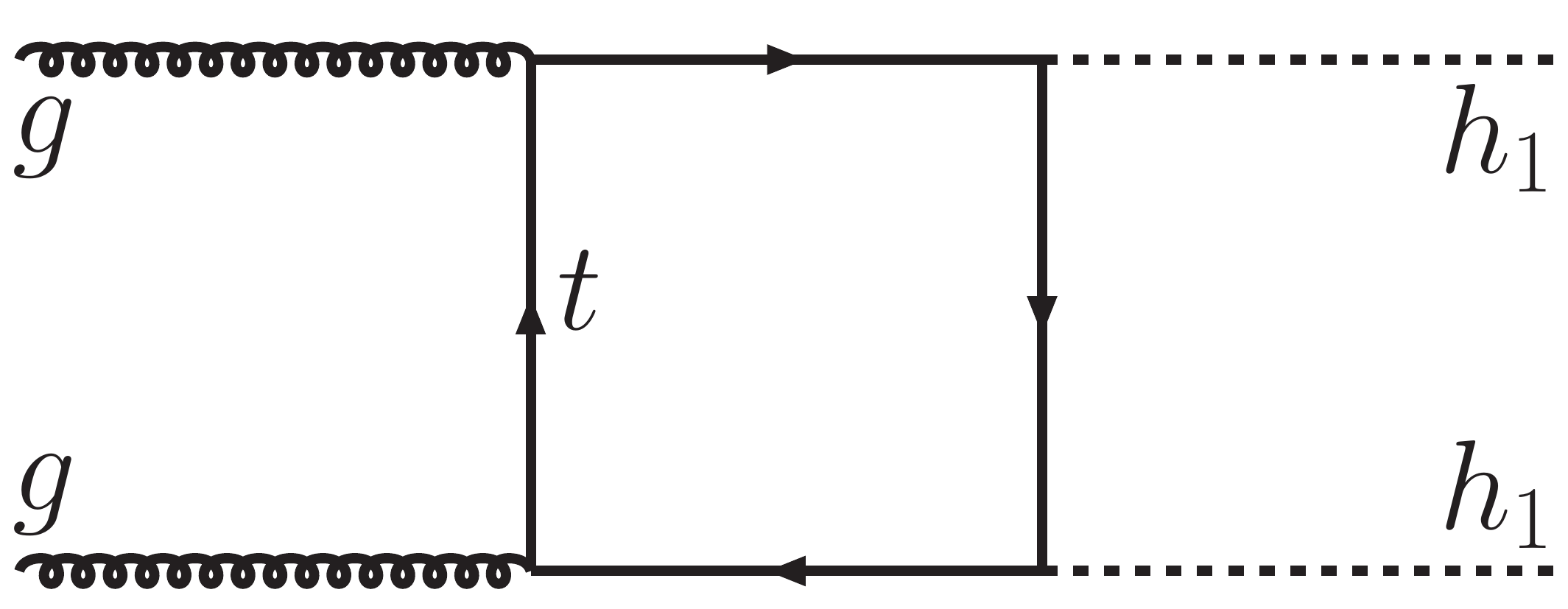}
  \caption{Representative resonant (left) and non-resonant (middle and right) Feynman diagrams contributing to di-Higgs production. 
  \label{fig:HHdiagram}}
\end{figure}
%%%%%%%%%%%%%%%%%%%%%%%%%%%%%%%%%%%%%%%%%%%%%%%%%%%%%%%%%%%%%%%%%%%%%%%%%%%%%%%%%%%%%%%%%%%%%%%%%%%%%%%%%%%%%%%%%

%%%%%%%%%%%%%%%%%%%%%%%%%%%%%%%%%%%%%%%%%%%%%%%%%%%%%%%%%%%%%%%%%%%%%%%%%%%%%%%%%%%%%%%%%%%%%%%%%%%%%%%%%%%%%%%%%
%%%%%%%%%%%%%%%%%%%%%%%%%%%%%%%%%%%%%%%%%%%%%%%%%%%%%%%%%%%%%%%%%%%%%%%%%%%%%%%%%%%%%%%%%%%%%%%%%%%%%%%%%%%%%%%%%
\begin{figure}[t]
  \centering
\includegraphics[width=0.32\textwidth]{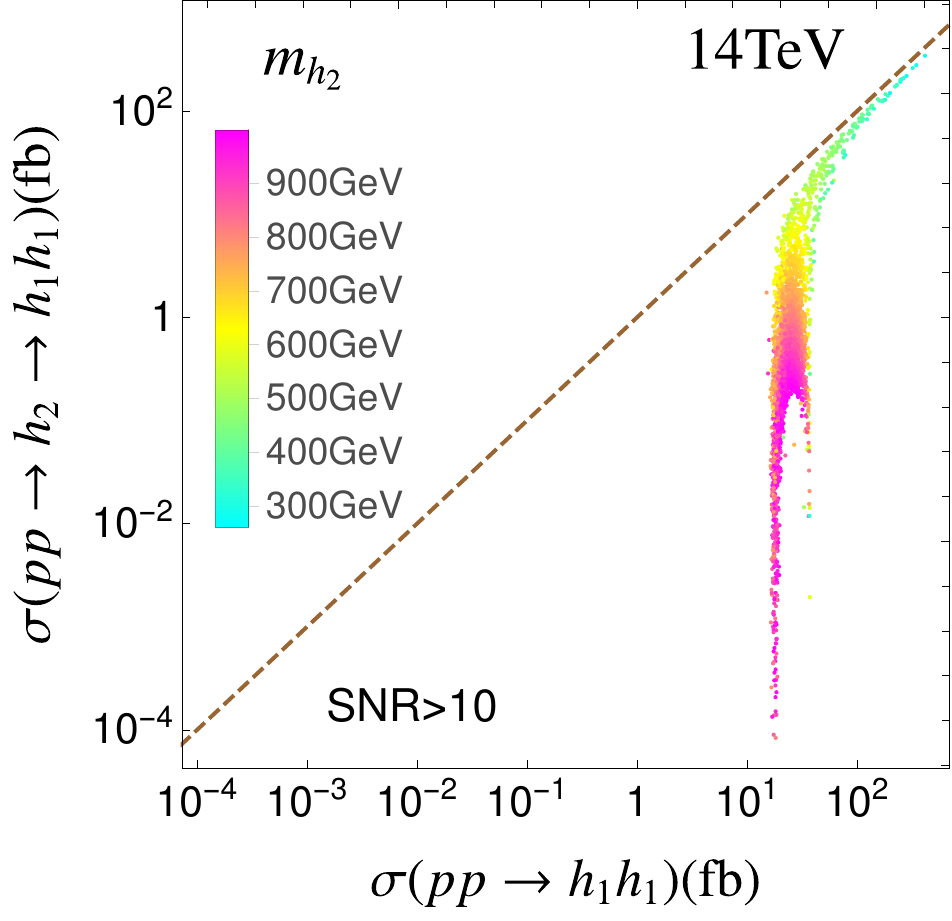}
\includegraphics[width=0.319\textwidth]{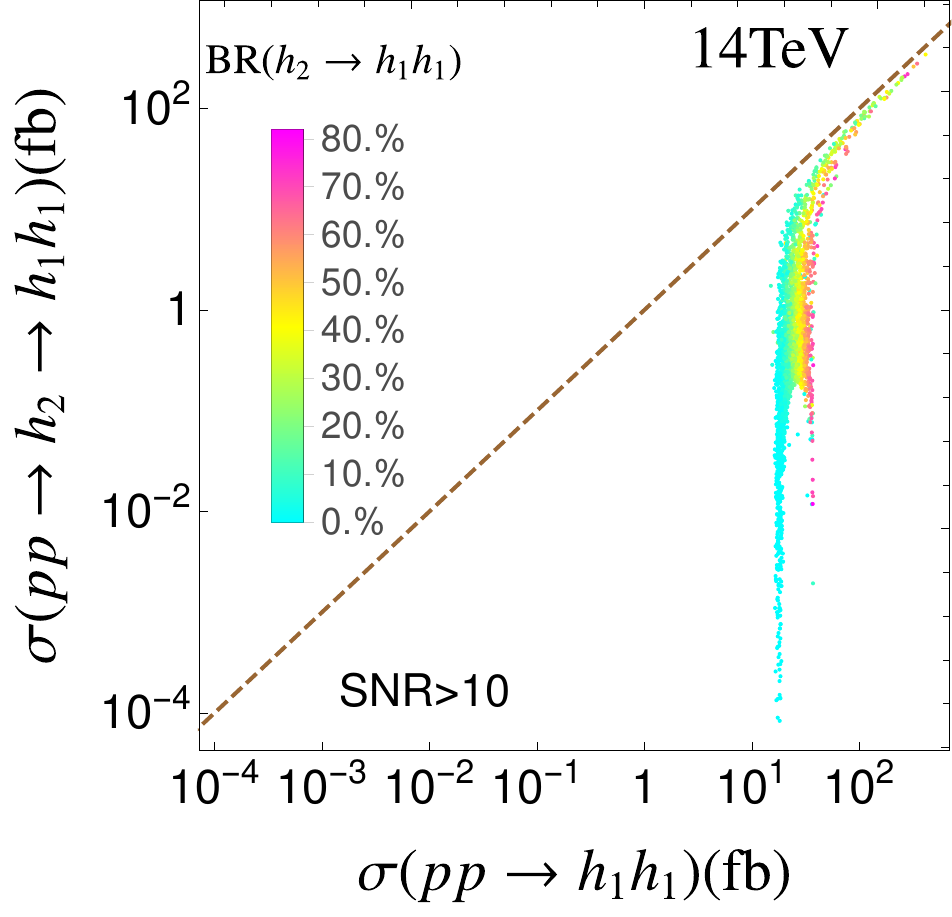}
\includegraphics[width=0.328\textwidth]{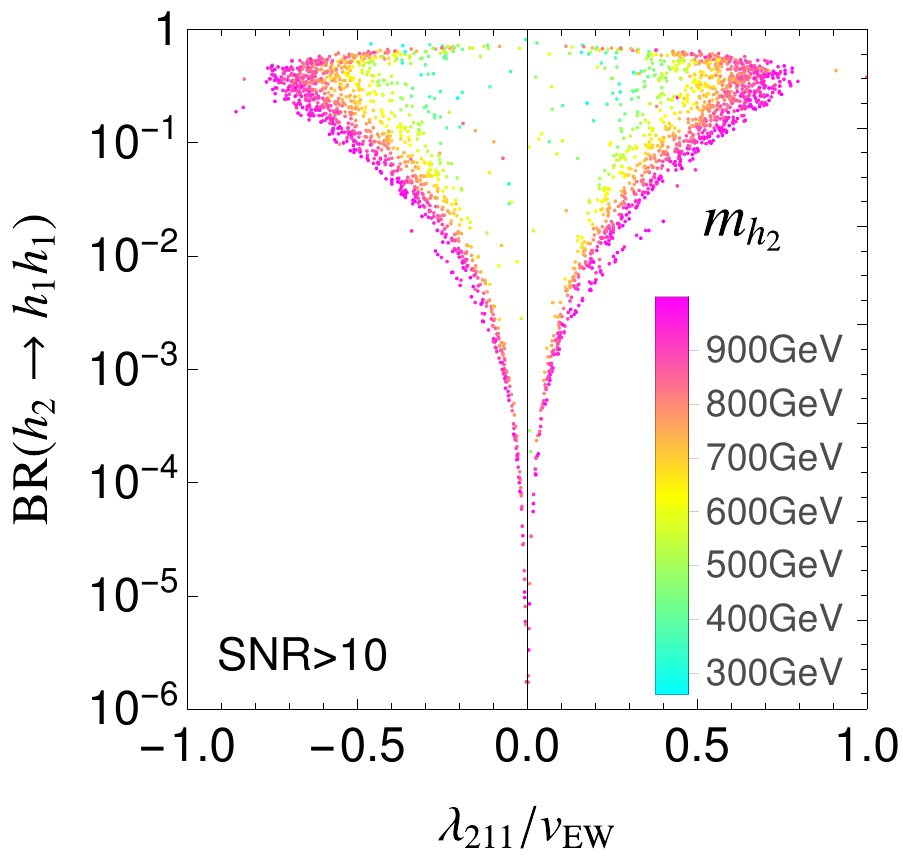}
\caption{
\label{fig:hh-resonant-full}
Resonant contribution to the cross section for di-Higgs production, versus the total cross-section.
The left plot shows the correlation of the two cross sections, with the colors denoting values of $m_{h_2}$.
The middle plot has the colors switched to the branching ratio of $h_2 \rightarrow h_1 h_1$. The right plot shows
this branching ratio versus the trilinear coupling $h_2 h_1 h_1$, where the color denotes $m_{h_2}$. In the left two plots,
the dashed line denotes the place where these two cross sections are the same.
}
\end{figure}
%%%%%%%%%%%%%%%%%%%%%%%%%%%%%%%%%%%%%%%%%%%%%%%%%%%%%%%%%%%%%%%%%%%%%%%%%%%%%%%%%%%%%%%%%%%%%%%%%%%%%%%%%%%%%%%%%
The leading order Feynman diagrams for double Higgs production occur at one-loop and consist of 
both the resonant and non-resonant channels, as shown in Fig.~\ref{fig:HHdiagram}. The non-resonant channel includes the box 
diagrams and a triangle diagram involving the vertex $h_1 h_1 h_1$.
The resonant channel is the production of a on-shell $h_2$ which subsequently decays into two Higgs, thus including the 
$h_2 h_1 h_1$ vertex.
The amplitude at leading order was given in the early papers~\cite{Eboli:1987dy,Plehn:1996wb} with the result expressed in terms of 
Passarino-Veltman scalar integrals. This result has also been implemented 
into~\texttt{MadGraph}~\cite{Alwall:2014hca} taking into account the presence 
of a heavier SM-like scalar~\footnote{https://cp3.irmp.ucl.ac.be/projects/madgraph/wiki/HiggsPairProduction}, 
which we use for calculating the corresponding cross sections for each point shown here. 
This takes as input the modified Higgs top Yukawa coupling, the Higgs trilinear coupling, the heavy scalar top 
coupling, the $h_2 h_1 h_1$ coupling and the mass as well as the decay width of $h_2$. 
Since $h_2$ decays into SM particles with reduced coupling $(-\sin\theta)$ as compared with the SM Higgs and also
decays to a pair of $h_1$, the total width is simply given by:
\begin{eqnarray}
\Gamma_{h_2} = \sin^2\theta \, \Gamma_{\text{SM}}(h_2 \rightarrow X_{SM}) + \Gamma(h_2 \rightarrow h_1 h_1) ,
\end{eqnarray}
where $\Gamma_{\text{SM}}(h_2 \rightarrow X_{SM})$ denotes an exact SM Higgs-like $h_2$ decaying into the SM particles.

For the di-Higgs production, if the resonant production of $h_1 h_1$ via the $h_2$ resonance dominates the cross section, then
the cross section can be written in the narrow width approximation as
\begin{eqnarray}
  \sigma(pp\rightarrow h_1 h_1) = \sigma(pp\rightarrow h_2) \text{BR}(h_2 \rightarrow h_1 h_1) .
\end{eqnarray}
In reality, interference effects between the resonant and non-resonant diagrams may be important and  
lead to constructive or destructive effect on the final full cross section~\cite{Carena:2018vpt}. We thus compare, for each scanned point, the
obtained cross section for both the full calculation and the above approximation from the purely  resonant
production. This is shown in the left and middle plots of Fig.~\ref{fig:hh-resonant-full} for 
$\sigma(pp\rightarrow h_1 h_1)$ versus $\sigma(p p \rightarrow h_2 \rightarrow h_1 h_1)$ for all the points which give 
detectable GW signals, that is, those with $\text{SNR}>10$. These cross sections are both calculated at leading order but we have 
added a common K-factor of $2.27$~\cite{deFlorian:2013uza} to take into account of higher order corrections.
The colors in the left panel denote the values of $m_{h_2}$ 
and those in the middle denote $\text{BR}(h_2 \rightarrow h_1 h_1)$. It is clear from these figures that the resonant 
cross section is always less than the full one-loop result and drops sharply  as $m_{h_2}$ is increased (left panel).
Since, as we have seen in previous sections, the points with large SNR are concentrated around the region with larger $m_{h_2}$, 
most of the points with detectable GWs turn out to give small di-Higgs production and even negligible resonant production.
The colors in the left panels make it clear that most of the points which have larger $m_{h_2}$ (and larger SNR) tend to give very small 
di-Higgs production, with a cross section of $\mathcal{O}(10)\text{fb}$, while smaller $m_{h_2}$ gives 
$\mathcal{O}(100)\text{fb}$. Moreover, there is a sharp drop of the resonant production cross section. From the middle panel, we
can see that the color of decreasing branching ratio $h_2 \rightarrow h_1 h_1$ coincides partly with increasing $m_{h_2}$ for 
the very large $m_{h_2}$ points. The small branching ratio is found for a majority of points and  is due to the 
smallness of $\lambda_{211}$. This can be seen from the right panel, where this correlation is shown with the color denoting
$m_{h_2}$. It is found that a majority of points which have large $m_{h_2}$ give small branching ratio. This can partly explain the cause of the drop of the resonant production. 
%%%%%%%%%%%%%%%%%%%%%%%%%%%%%%%%%%%%%%%%%%%%%%%%%%%%%%%%%%%%%%%%%%%%%%%%%%%%%%%%%%%%%%%%%%%%%%%%%%%%%%%%%%%%%%%%%
%%%%%%%%%%%%%%%%%%%%%%%%%%%%%%%%%%%%%%%%%%%%%%%%%%%%%%%%%%%%%%%%%%%%%%%%%%%%%%%%%%%%%%%%%%%%%%%%%%%%%%%%%%%%%%%%%
\begin{figure}[t]
\centering
\includegraphics[width=0.6\textwidth]{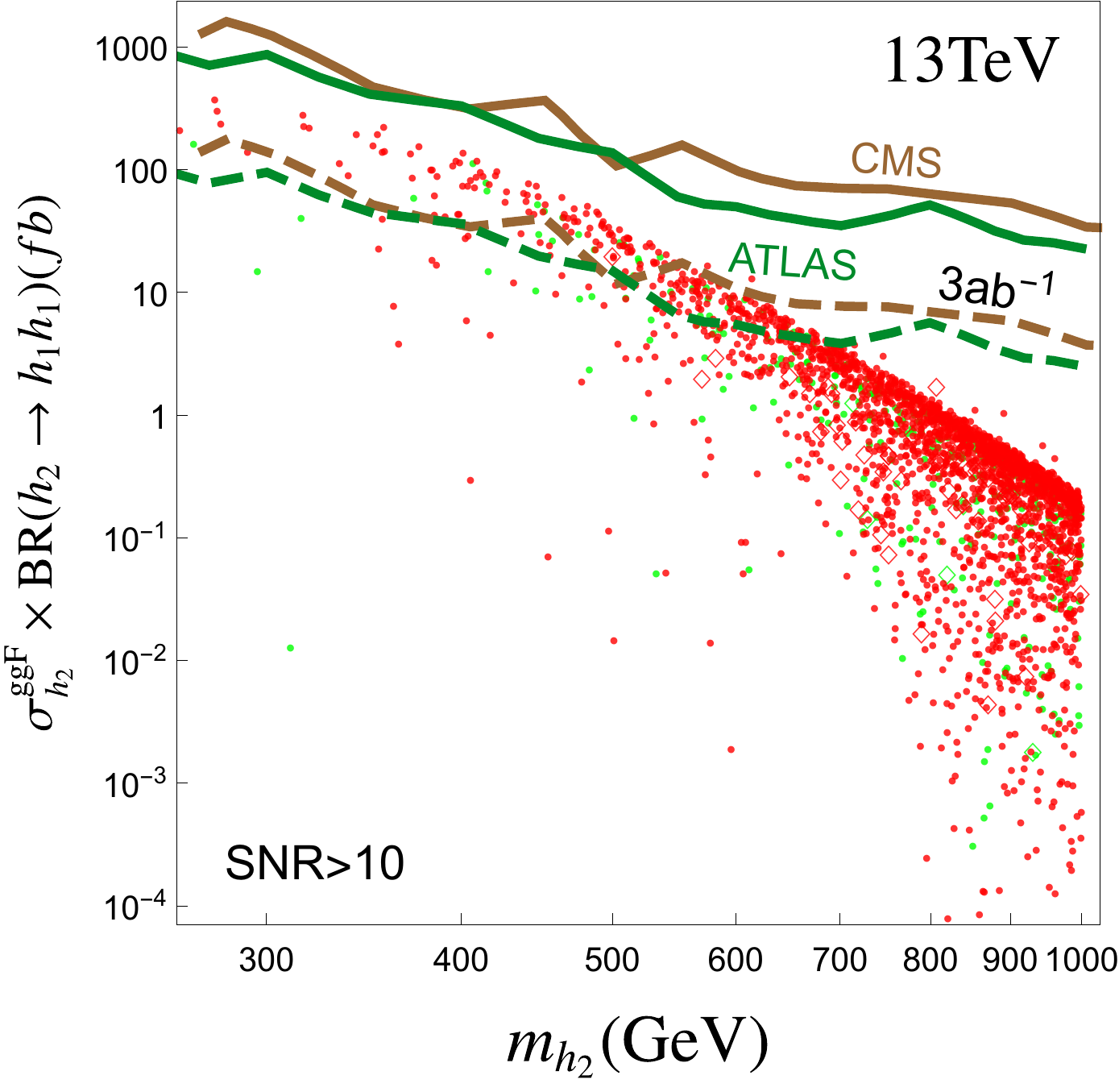}
\caption{\label{fig:HH}
The upper limits on di-Higgs resonant production cross section from ATLAS and CMS combined searches,
shown as solid green and brown lines for ATLAS and CMS, respectively. The dashed 
lines denote the corresponding future projections for $3 \text{ab}^{-1}$ of data at the  HL-LHC (13TeV).
As in the other plots, we distinguish those points which give SNR $>$ 50 (red) and those of
50 $>$ SNR $>$ 10 (green).
}
\end{figure}
%%%%%%%%%%%%%%%%%%%%%%%%%%%%%%%%%%%%%%%%%%%%%%%%%%%%%%%%%%%%%%%%%%%%%%%%%%%%%%%%%%%%%%%%%%%%%%%%%%%%%%%%%%%%%%%%%
%%%%%%%%%%%%%%%%%%%%%%%%%%%%%%%%%%%%%%%%%%%%%%%%%%%%%%%%%%%%%%%%%%%%%%%%%%%%%%%%%%%%%%%%%%%%%%%%%%%%%%%%%%%%%%%%%

On the experimental side, both the ATLAS and CMS collaborations have recently published their search results for 
non-resonant and resonant di-Higgs productions using the data collected in 2016 at 13 TeV, with nearly the same 
integrated luminosity. The CMS search result is based on the $35.9\text{fb}^{-1}$ data, in the di-Higgs 
decay channels $b\bar{b}\gamma\gamma$~\cite{Sirunyan:2018iwt}, 
$b\bar{b}\tau^+\tau^-$~\cite{Sirunyan:2017djm}, 
$b\bar{b}b\bar{b}$~\cite{Sirunyan:2017isc,Sirunyan:2018zkk,Sirunyan:2018qca,Sirunyan:2018tki}
and $b\bar{b}WW/ZZ$~\cite{Sirunyan:2017guj}, with a recent combination given in~\cite{Sirunyan:2018two}.
ATLAS used $36.1\text{fb}^{-1}$ data and searched in channels
$\gamma\gamma b \bar{b}$~\cite{Aaboud:2018ftw}, $b \bar{b} \tau^+ \tau^-$~\cite{Aaboud:2018sfw},
$b\bar{b}b\bar{b}$~\cite{Aaboud:2018knk}, $W W^{(\ast)} W W^{(\ast)}$~\cite{Aaboud:2018ksn} and 
$b\bar{b}W W^{\ast}$~\cite{Aaboud:2018zhh}, with also a combination of the first three channels~\cite{ATLAS:2018otd}. 
We use the ATLAS and CMS combined limits in the resonant production channels 
and show them with green and brown solid lines respectively in Fig.~\ref{fig:HH}. 
For the points giving detectable GWs, we calculate the resonant cross sections from gluon fusion at 
NNLO+NNLL using the available result in Ref.~\cite{deFlorian:2016spz}.
We can see that none of the points with detectable GW gives cross section above this limit. 
With the anticipation of HL-LHC at a luminosity of $3 \text{ab}^{-1}$ ($13\text{TeV}$), 
we can get the future projections of this limit by a simple rescaling and obtain the two dashed lines. 
For this projection, the region with lower $m_{h_2}\lesssim 550\text{GeV}$ can be partly explored by CMS 
and a little bit higher for ATLAS, while the high mass region remains out of reach for di-Higgs searches.
Yet, Some points of the scanned parameters space with observable SNR show a promising di-Higgs production cross section
of 50 fb or more at the LHC which, in principle, can be probed with 3 ab$^{-1}$.
Therefore GW measurements can complement collider searches by revealing the high $m_{h_2}$ region of the 
xSM model.

\subsection{Higgs Cubic and Quartic Couplings}

Future precise measurements of the Higgs cubic and quartic self-couplings can be used 
to reconstruct the 
Higgs potential to confirm ultimately the mechanism of EW symmetry breaking \footnote{The Lorentz structure of $hWW$ coupling already gave us some insight about the nature of EW symmetry breaking at the leading order.} and shed light on the nature of the EWPT. 
The measurements of above double Higgs production can be used to determine the cubic coupling and there
have been extensive studies on this topic~\cite{DiVita:2017vrr,Adhikary:2017jtu,Banerjee:2018yxy}. The best sensitivities obtained for these 
future colliders is typically at $\mathcal{O}(1)$.
Despite the more formidable challenges with the quartic coupling measurement, there is now growing 
interest in it. Several different methods have been proposed and studied: through triple Higgs 
production measurement~\cite{Plehn:2005nk},
through double Higgs production at hadron colliders where the quartic coupling enters 
$gg\rightarrow h h$ at two-loop~\cite{Bizon:2018syu}
or renormalizes the cubic coupling, and at lepton 
colliders(via Z-associated production $e^+ e^- \rightarrow Z hh$ and VBF production 
$e^+e^-\rightarrow \nu \nu h h$), 
where the quartic coupling is involved in the $VVhh$ coupling at one loop~\cite{Liu:2018peg}.
For example, Ref.~\cite{Liu:2018peg} found a precision of measurement of $\sim \pm 25$ for 
($500\text{GeV}, 4\text{ab}^{-1}$ + 1 TeV, $2.5 \text{ab}^{-1}$) and 
$\sim \pm 20$  for ($500\text{GeV}, 4\text{ab}^{-1}$ + 1 TeV, $8 \text{ab}^{-1}$) at
$1\sigma\text{C.L.}$, when the cubic coupling is marginalized in their $\chi^2$ analysis.

%%%%%%%%%%%%%%%%%%%%%%%%%%%%%%%%%%%%%%%%%%%%%%%%%%%%%%%%%%%%%%%%%%%%%%%%%%%%%%%%%%%%%%%%%%%%%%%%%%%%%%%%%%%%%%%%%
%%%%%%%%%%%%%%%%%%%%%%%%%%%%%%%%%%%%%%%%%%%%%%%%%%%%%%%%%%%%%%%%%%%%%%%%%%%%%%%%%%%%%%%%%%%%%%%%%%%%%%%%%%%%%%%%%
\begin{figure*}[t]
\includegraphics[width=\textwidth]{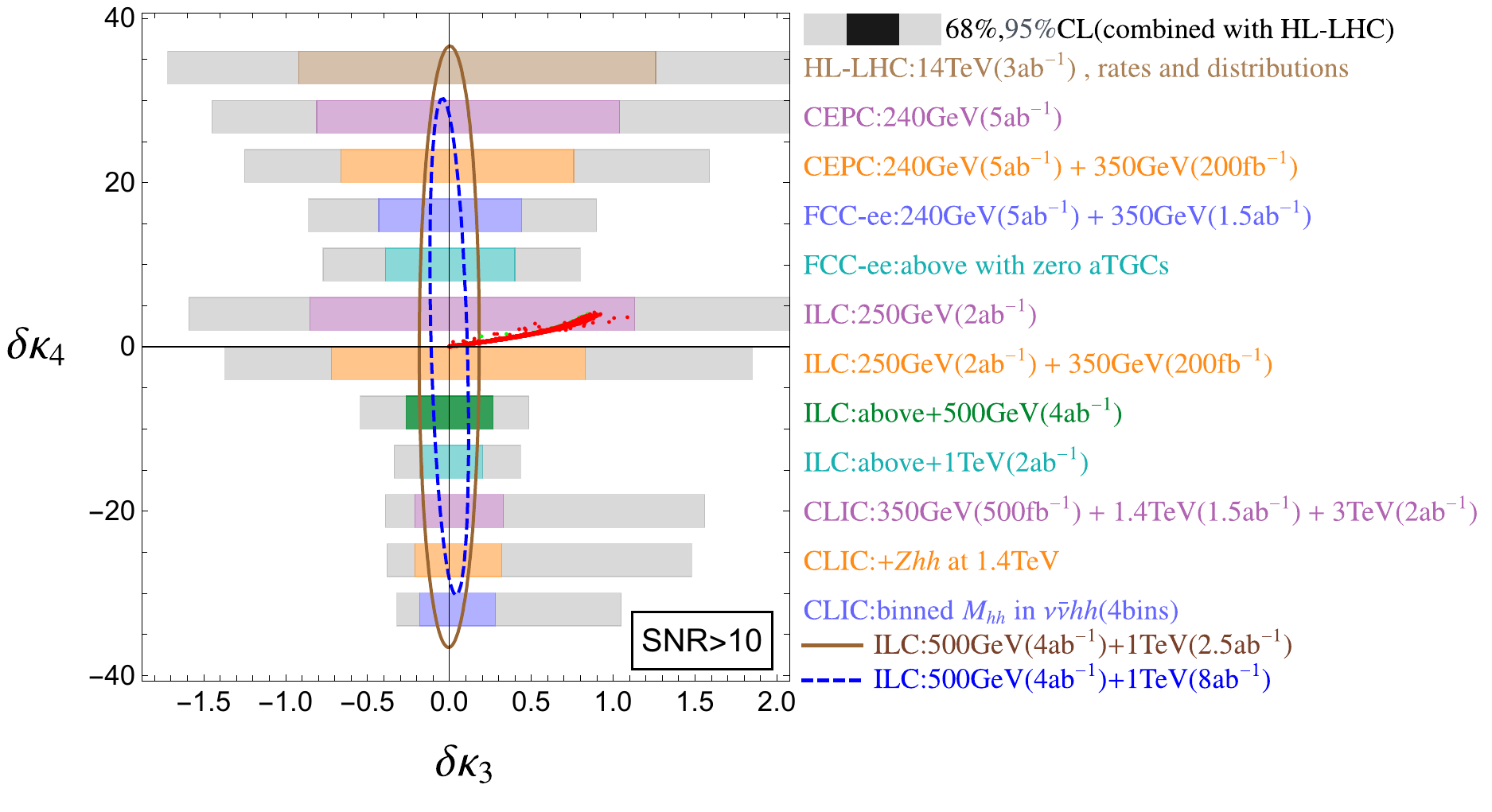}
\\
\includegraphics[width=\textwidth]{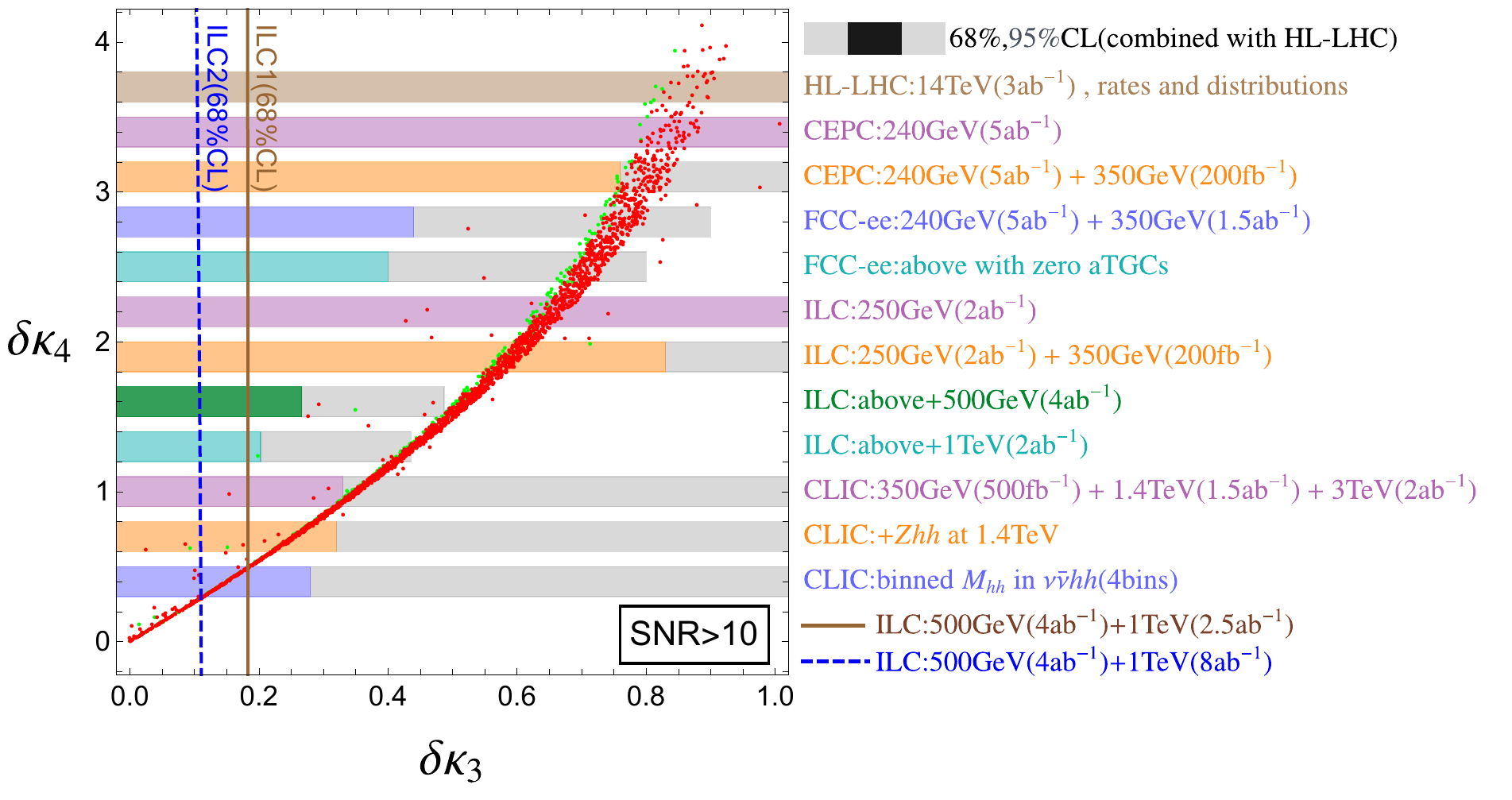}
\caption{\label{fig:k3k4}
The Higgs cubic and quartic couplings $(\Delta \kappa_3, \Delta \kappa_4)$ for parameter space points giving detectable GW.
Here the green points give $\text{SNR}>10$ and the red gives $\text{SNR}>50$.
The bars denote the sensitivity of $\Delta \kappa_3$ from a global analysis of future colliders in 
Ref.~\cite{DiVita:2017vrr}, for various detector scenarios shown on the right side of the figures.
The brown solid and blue dashed 
lines are the $1\sigma$ contours for two different ILC scenarios taken from Ref.~\cite{Liu:2018peg}.
The bottom panel is a zoomed-in version of the top one.
}
\end{figure*}
%%%%%%%%%%%%%%%%%%%%%%%%%%%%%%%%%%%%%%%%%%%%%%%%%%%%%%%%%%%%%%%%%%%%%%%%%%%%%%%%%%%%%%%%%%%%%%%%%%%%%%%%%%%%%%%%%
%%%%%%%%%%%%%%%%%%%%%%%%%%%%%%%%%%%%%%%%%%%%%%%%%%%%%%%%%%%%%%%%%%%%%%%%%%%%%%%%%%%%%%%%%%%%%%%%%%%%%%%%%%%%%%%%%
In the xSM, both the Higgs cubic and quartic couplings are modified compared with their SM counterparts:
\begin{eqnarray}
  &&  i \lambda_{h_1 h_1 h_1} = 
  6 \Big[
  \lambda  v c_{\theta }^3 + 
\frac{1}{4} c_{\theta }^2 s_{\theta } \left(2 a_2 v_s+a_1\right)+\frac{1}{2} a_2 v c_{\theta }
   s_{\theta }^2 \nonumber \\
   && \hspace{1.6cm} +\frac{1}{3} s_{\theta }^3 \left(3 b_4 v_s+b_3\right) \Big] , \\
 && i \lambda_{h_1 h_1 h_1 h_1} = 6(\lambda c_{\theta}^4 + a_2 s_{\theta}^2 c_{\theta}^2 + b_4 s_{\theta}^4).
\end{eqnarray}
In the absence of mixing of the scalars($\theta = 0$), these couplings reduce to the 
corresponding SM values $i \lambda_{h_1 h_1 h_1} = 3 \mha^2/v$ 
and $i \lambda_{h_1 h_1 h_1 h_1}=3\mha^2/v^2$. When $\theta \neq 0$, 
we parametrize the deviations of these couplings from the SM values as:
\begin{eqnarray}
  \Delta \mathcal{L} = - \frac{1}{2}\frac{m_{h_1}^2}{v} (1 + \delta \kappa_3) h_1^3 - 
  \frac{1}{8} \frac{m_{h_1}^2}{v^2} (1 + \delta \kappa_4) h_1^4 ,
\end{eqnarray}
and show in Fig.~\ref{fig:k3k4} these values for the points that give detectable GWs. 
The features that we can read from this figure are:(1) both $\delta \kappa_3$ and $\delta \kappa_4$ are positive;
(2) both variations are $\mathcal{O}(1)$ as $\delta \kappa_3 \in (0,1)$ and $\delta \kappa_4 \in (0,4)$;
(3) a correlation exists $\delta \kappa_4 \equiv \eta \delta \kappa_3$, with 
$\eta \approx 2.8$ for $\delta \kappa_3 \lesssim 0.4$ and most points fall within $\eta \in (2,4)$.
To understand these, we note, since phenomenological constraints requires a small $\theta$,
we expect the second feature to follow naturally. The other features can be understood by Taylor expanding
the couplings for small $\theta$ and we find:
\begin{eqnarray}
&& \delta \kappa_3 = \theta^2 \left[-\frac{3}{2} + \frac{2 m_{h_2}^2 -2 b_3 v_s - 4 b_4 v_s^2}{m_{h_1}^2}\right] + \mathcal{O}(\theta^3), \nonumber \\
&& \delta \kappa_4 = \theta^2 \left[-3 + \frac{5 m_{h_2}^2 - 4 b_3 v_s - 8 b_4 v_s^2}{m_{h_1}^2}\right] + \mathcal{O}(\theta^3)  . \quad
\end{eqnarray}
In the above square brackets, the terms proportional to $m_{h_2}^2/m_{h_1}^2$ dominate for the majority of the points since 
$v_s$ is concentrated at small values; $b_3$ is at most $\sim 10 v_{\text{EW}}$, $b_4 \lesssim 5$ from the scan and 
$m_{h_2} \gtrsim 500\text{GeV}$ generally holds. Then the above approximations show positive $\delta \kappa_3$ and $\delta \kappa_4$
and give $\delta \kappa_4/\delta \kappa_3 \approx 2.5$, which is fairly close to $\eta = 2.8$. 
For relatively large $\theta$, high order corrections need to be taken into account and above linear 
correlation would be changed.

To compare with the direct measurements of these couplings at  future $e^+e^-$ colliders and the HL-LHC, we added in 
Fig.~\ref{fig:k3k4} the precisions of these measurements from studies in the literature. 
The two elliptical $68\%$CL closed contours are taken from Ref.~\cite{Liu:2018peg} which focuses on the 
quartic coupling, for two possible scenarios of the ILC. 
The bars are the precisions that can be reached from various considerations of future colliders, labelled
on the right of the figure, taken from Ref.~\cite{DiVita:2017vrr}(for other studies, see 
e.g.~\cite{Borowka:2018pxx,Bizon:2018syu,Kilian:2018bhs,Jurciukonis:2018skr,Borowka:2018pxx,Maltoni:2018ttu,Adhikary:2017jtu,Banerjee:2018yxy}). Here the inner and outer bar regions denote the $68\%$CL and $95\%$CL results.
We can see, it is generically very hard for colliders to probe the cubic coupling at a precision 
that can reveal the points giving detectable GWs with high confidence level(say 95\%)~\footnote{
It should be noted that both studies used some versions of the effective field theory approach to quantify the modification of 
the SM couplings due to possible new physics effects. Therefore the precisions overlaid in Fig.~\ref{fig:k3k4} might not be what
the colliders can achieve if the xSM model was used in their studies. However we expect the two contours, taken from Ref.~\cite{Liu:2018peg},
to be largely unaffected since the heavier scalar contribution in their framework is suppressed by extra powers of $s_{\theta}$. 
We also expect that the bar regions, taken from Ref.~\cite{DiVita:2017vrr}, would get tighter since the set of parameters used in their study 
are highly correlated here and the resonant contribution was not included in their analyses.
}.
The most precise comes from the ILC when all possible runs at different luminosities 
are combined and with the data of HL-ILC included, which gives $0.4 \sim 0.5$ uncertainty on the 
measurement of $\delta \kappa_3$ at $95\%$CL. While the analysis in Ref.~\cite{DiVita:2017vrr} does not 
include the quartic coupling, the contours from Ref.~\cite{Liu:2018peg} do 
give a hint on its measurement and show
that it is infeasible for the colliders to probe the parameter space giving detectable GWs. For the trilinear and quartic coupling deviations
that we found, the impact on the triple Higgs cross section is mild
for hadron colliders even for a future $pp$ collider at 100 TeV~\cite{Plehn:2005nk,Binoth:2006ym}, however,
resonant contributions in xSM might enhance the cross section up to a factor of ${\cal O}(10)$~\cite{Chen:2015gva}.

Therefore we expect future GW measurements can make a valuable complementary role in determining the Higgs
self-couplings, especially the quartic coupling. While we do not have a statistical analysis here,
Fig.~\ref{fig:k3k4} does tell us that $\delta \kappa_4$ is equally important as $\delta \kappa_3$ on GW 
signal generation since $\eta$ is at most $4$. Thus we expect a full
statistical analysis would yield roughly the same precision on the determination of $\delta \kappa_3$ and
$\delta \kappa_4$, which is well improved compared with the situation at colliders.

%Moreover, we also find points where the quartic couplings of the
%SM Higgs boson deviate significantly from the SM and that should also be observable at future GW experiments. The modified triple and quartic couplings
%might conspire~\cite{Plehn:2005nk,Binoth:2006ym} to enhance triple Higgs production at colliders somewhat.

%%%%%%%%%%%%%%%%%%%%%%%%%%%%%%%%%%%%%%%%%%%%%%%%%%%%%%%%%%%%%%%%%%%%%%%%%%%%%%%%%%%%%%%%%%%%%%%%%%%%%%%%%%%%%%%%%
\begin{figure}[t]
  \centering
\includegraphics[width=0.6\textwidth]{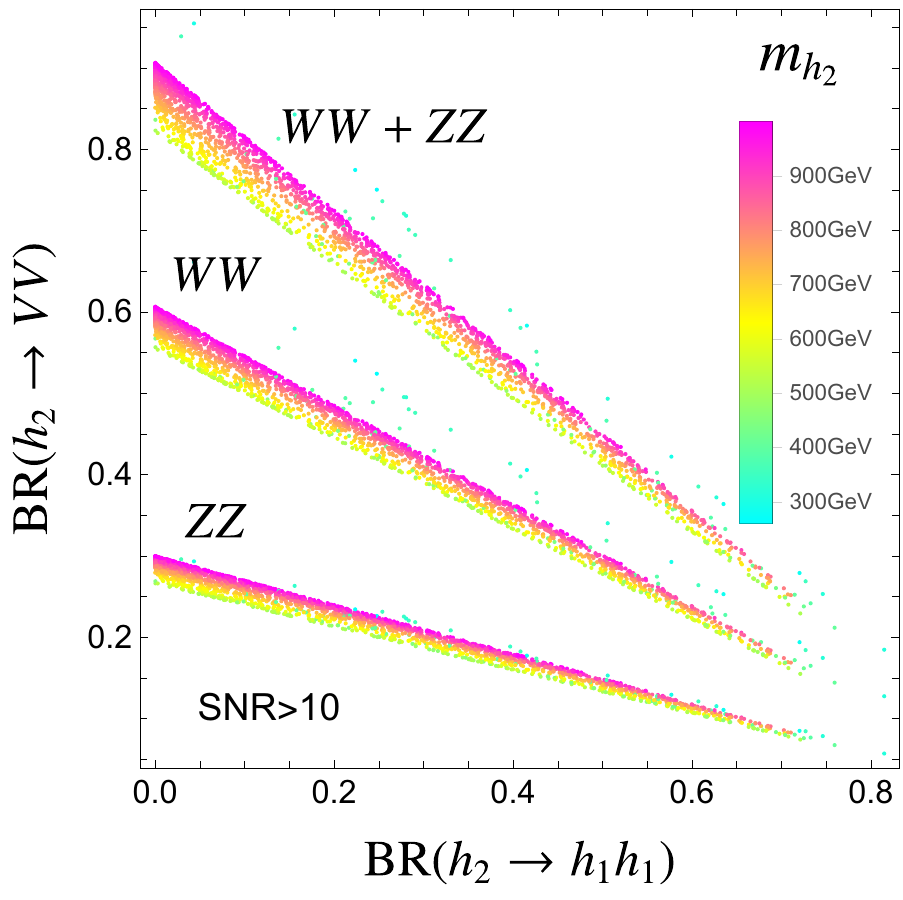}
\caption{\label{fig:brvv}
The branching ratios of $h_2$ in $h_1 h_1$ and $VV$ final states, where $VV=WW,ZZ,WW+ZZ$, with 
the color denoting the value of $m_{h_2}$.
}
\end{figure}
%%%%%%%%%%%%%%%%%%%%%%%%%%%%%%%%%%%%%%%%%%%%%%%%%%%%%%%%%%%%%%%%%%%%%%%%%%%%%%%%%%%%%%%%%%%%%%%%%%%%%%%%%%%%%%%%%
\subsection{Diboson Resonance Search Limits at Colliders}

The $WW$ and $ZZ$ branching ratios
become sizeable in parts of the parameter space where the trilinear coupling $\lambda_{211}$ is relatively small, as one can see from the
rightmost panel of Fig.~\ref{fig:hh-resonant-full}. In Fig.~\ref{fig:brvv}, we show the branching ratios of the $h_2\rightarrow WW,ZZ$ and $h_2\rightarrow h_1h_1$ channels. We see that the $WW,ZZ$ channels can be as big as 90\% for a large range of $h_2$ masses which could show up at searches for weak diboson resonances. Combined, $WW,ZZ$ and $h_1h_1$ correspond to nearly  all the decays of  $h_2$, which make them the best search channels for $h_2$ resonances at colliders.

\begin{figure}[t]
  \centering
\includegraphics[width=0.45\textwidth]{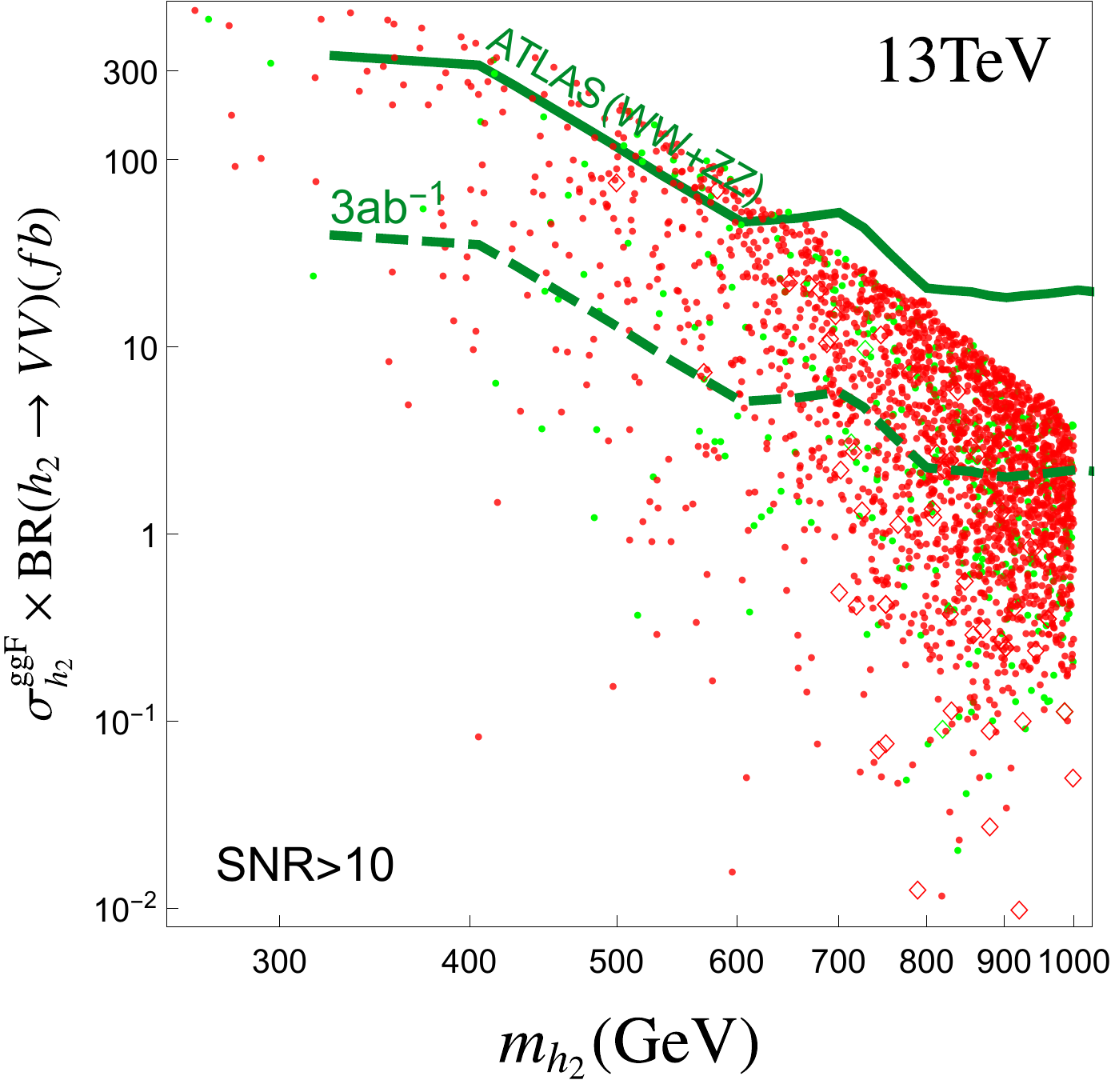}
\quad
\includegraphics[width=0.45\textwidth]{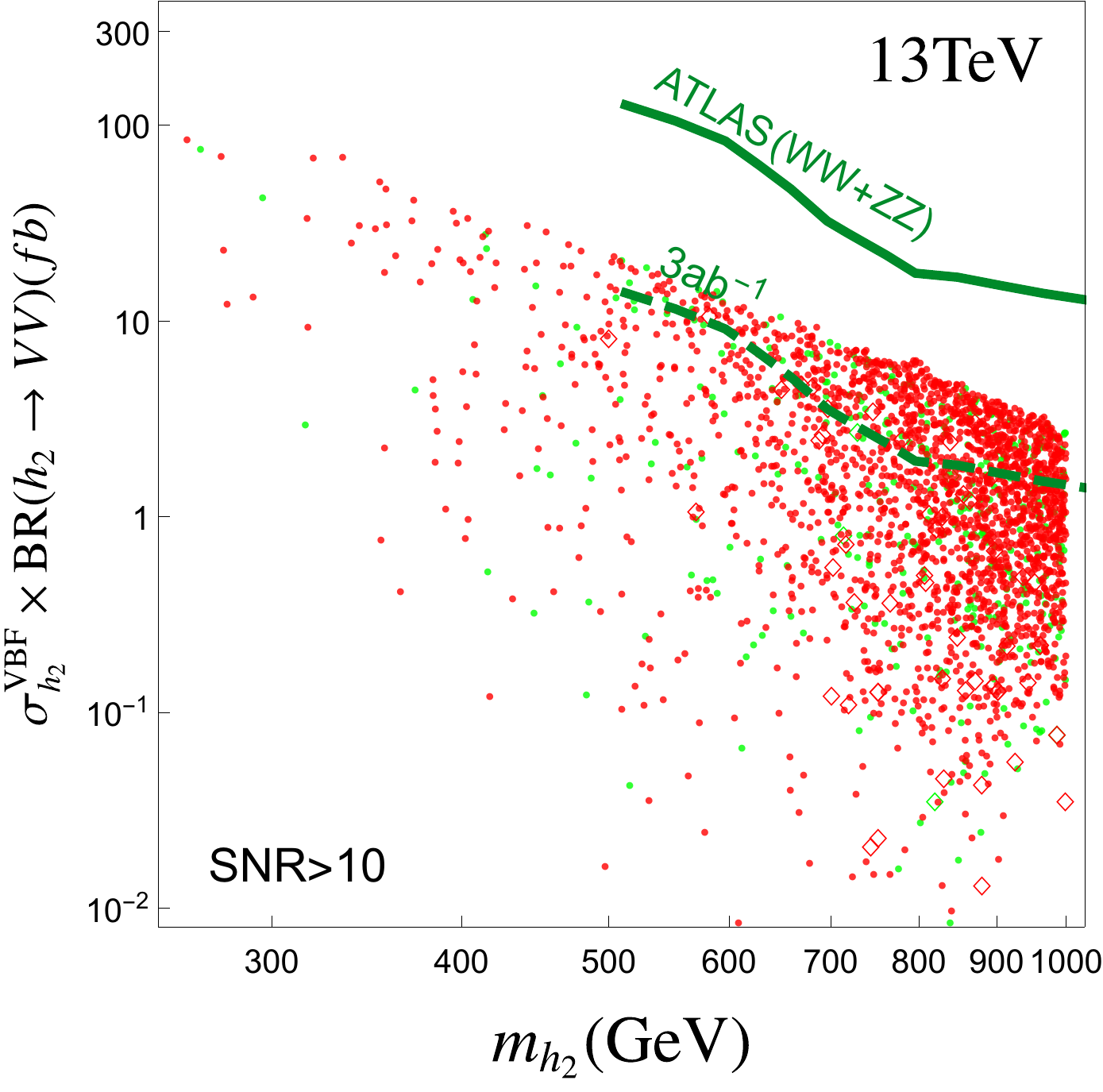}
\caption{\label{fig:LHC-mh2} 
Combined limits from ATLAS (solid line) and  future HL-LHC projections (dashed line) for searches of a heavy SM-like resonance in 
the $WW/ZZ$ channel from gluon fusion (left) and vector boson fusion production (right). 
As in the other plots, we distinguish those points which give SNR $>$ 50(red) and those of
50 $>$ SNR $>$ 10(green).
}
\end{figure}

Besides the di-Higgs production measurements, which can be used to extract the Higgs cubic and 
quartic couplings, there also exist generic scalar resonance searches at the LHC.
In particular, ATLAS and CMS have performed extensive analyses in the searches for a 
heavier SM-like scalar resonance in $VV$ and $VH$ decay channels of the heavy scalar ($V=W/Z$). 
ATLAS gives a recent combination of all
previous analyses in bosonic and leptonic final states at $\sqrt{s}=13\text{TeV}$ with $36\text{fb}^{-1}$ data 
collected in 2015 and 2016~\cite{Aaboud:2018bun}. The limits are drawn for $h_2$ production cross section in 
gluon fusion and vector boson fusion production channels. These two limits are shown in the left and right panels, respectively, 
in Fig.~\ref{fig:LHC-mh2} with green solid lines, together with the detectable GW points. 
For cross section calculations, we use the set of result calculated to NNLO precision for VBF and 
for gluon fusion, we use NNLO+NNLL, as also used before in Fig.~\ref{fig:HH}.
%%%%%%%%%%%%%%%%%%%%%%%%%%%%%%%%%%%%%%%%%%%%%%%%%%%%%%%%%%%%%%%%%%%%%%%%%%%%%%%%%%%%%%%%%%%%%%%%%%%%%%%%%%%%%%%%%
%%%%%%%%%%%%%%%%%%%%%%%%%%%%%%%%%%%%%%%%%%%%%%%%%%%%%%%%%%%%%%%%%%%%%%%%%%%%%%%%%%%%%%%%%%%%%%%%%%%%%%%%%%%%%%%%%
%%%%%%%%%%%%%%%%%%%%%%%%%%%%%%%%%%%%%%%%%%%%%%%%%%%%%%%%%%%%%%%%%%%%%%%%%%%%%%%%%%%%%%%%%%%%%%%%%%%%%%%%%%%%%%%%%
%%%%%%%%%%%%%%%%%%%%%%%%%%%%%%%%%%%%%%%%%%%%%%%%%%%%%%%%%%%%%%%%%%%%%%%%%%%%%%%%%%%%%%%%%%%%%%%%%%%%%%%%%%%%%%%%%

It is evident that the current limits from diboson searches are rather loose as most points fall 
under this line, with gluon fusion limit being able to touch a fraction of the lighter $h_2$ point.
For the HL-LHC with $\sim 3 \text{ab}^{-1}$, 
we obtain estimates of future projections by a simple scaling factor and obtain
the dashed lines for $\sim 3 \text{ab}^{-1}$ at $13\text{TeV}$ (while HL-LHC would probably run at $14\text{TeV}$).
We can see in all cases that the HL-LHC will probe a larger fraction of the parameter space for both ggH and
VBF channels. For ggH, this region covers a range from low to high masses. For VBF, it can cover a region of relatively 
heavy $h_2$. Both channels are sensitive to $h_1h_1$ cross section times branching ratio down to $\sim 1$ fb in some favorable
points of the parameters space. The points that can be probed by HL-LHC serve as promising targets for 
both colliders and GW detectors but a majority of the parameter space will probably be left to GW detectors.

\section{\label{sec:summary}Summary}

In this paper, we embarked on a  study of the singlet-extended SM Higgs sector. A detailed scan of the parameter space of this model was performed, incorporating all relevant phenomenological constraints, and regions with large SNR at LISA were identified. Subtle issues pertaining to the bubble wall velocity were discussed, and a range of velocity profiles described. 

Our main findings are the following.  
For the parameter space that satisfies all phenomenological constraints, gives successful EWPT and 
generates GWs, $99\%$ leads to a one-step EWPT with the remaining to two-step EWPT and $22\%$ generates
detectable GWs($\text{SNR}>10$) at LISA. The main features of the parameter space that gives 
detectable GWs is: $20 \text{GeV}\lesssim |v_s| \lesssim 50\text{GeV}$, where $v_s$ is the vev of 
the singlet field; it is more concentrated in the large $m_{h_2}$ region, where $m_{h_2}$ is the 
mass of the heavier scalar $h_2$; $\theta \lesssim 0.2$ for the majority of the space. 
Di-Higgs searches at both ATLAS and CMS are currently unable to probe this parameter space, but HL-LHC will be able 
to probe the lighter $h_2$ region while the heavier $h_2$ region will remain elusive. Weak diboson resonance searches cannot constrain xSM much either but the HL-LHC will be able to probe a large fraction of its parameters space in this channel. The Higgs cubic and quartic couplings are at $\mathcal{O}(1)$ deviations from the SM values and obey a relation $\delta \kappa_4 \approx (2-4) \delta \kappa_3$, where $\delta \kappa_4$ and $\delta \kappa_3$ are  the relative deviations of the quartic and cubic couplings from their SM counterparts respectively.

Our results broadly indicate that high energy colliders and GW detectors are going to play complementary roles in probing the parameter space of scalar sectors. Several future directions can be contemplated. It would be interesting to understand how this complementarity plays out in two Higgs doublet models, as well as other scalar sector extensions classified in \cite{Chung:2012vg}. It would also be interesting to investigate the complementarity of GW and collider probes for phase transitions in the dark sector. We leave these questions for future study.

\section{Acknowledgments}
A. Alves thanks Conselho Nacional de Desenvolvimento Cient\'ifico (CNPq) for its financial support,
grant 307265/2017-0. K. Sinha is supported by the U. S. Department of Energy grant desc0009956. T. Ghosh is supported by U. S. Department of Energy grant de-sc0010504. We would like to thank David Curtin, Ian Lewis, Hao-Lin Li and Ligong Bian for helpful discussions. We also thank Yi-Ming Hu and Zheng-Cheng Liang of the
Tianqin program for sending us the Tianqin sensitivity curve and for valuable discussions.

\appendix

\section{\label{sec:unitarity}Perturbative Unitarity S Matrix}
We consider a total of eleven $2\rightarrow 2$ channels of scalars and longitudinal gauge bosons scatterings.
These are grouped into seven charge neutral channels $(h_1 h_1, h_2 h_2, h_1 h_2,  h_1 Z, h_2 Z, Z Z, W^+ W^-)$, 
three charge-1 channels $(h_1 W^+, h_2 W^+, Z W^+)$ and one charge-2 channel $(W^+ W^-)$.
The leading partial wave amplitudes of these scatterings are given collectively by a symmetric matrix, 
which itself is a direct sum of the matrices from these three groups:
$\mathcal{S} = \mathcal{S}_0 \bigoplus \mathcal{S}_1 \bigoplus \mathcal{S}_2$.
The tree level perturbative unitarity requires that the absolute value of each eigenvalue of this matrix is less than $(1/2 \times 16\pi)$.
The non-zero elements of the $7\times 7$ matrix $\mathcal{S}_0$ is listed as follows(see e.g., Ref.~\cite{Kanemura:2015ska} for a detailed calculation):
\begin{eqnarray}
&& \mathcal{S}_{11} = -3 \left(a_2 c_{\theta }^2 s_{\theta }^2+b_4 s_{\theta }^4+\lambda  c_{\theta }^4\right), \nonumber  \\
&& \mathcal{S}_{12} = \frac{1}{8} \left(3 \cos (4 \theta ) \left(-a_2+b_4+\lambda \right)-a_2-3 b_4-3 \lambda \right), \nonumber\\
&& \mathcal{S}_{13} = \frac{3 \sin (2 \theta ) \left(\cos (2 \theta ) \left(-a_2+b_4+\lambda \right)-b_4+\lambda \right)}{2 \sqrt{2}}, \nonumber\\
&& \mathcal{S}_{16} = -\frac{1}{2} a_2 s_{\theta }^2-\lambda  c_{\theta }^2,\nonumber \\
&& \mathcal{S}_{17} = -\frac{a_2 s_{\theta }^2+2 \lambda  c_{\theta }^2}{\sqrt{2}}, \nonumber \\
&& \mathcal{S}_{22} = -3 \left(a_2 c_{\theta }^2 s_{\theta }^2+b_4 c_{\theta }^4+\lambda  s_{\theta }^4\right),\nonumber \\
&& \mathcal{S}_{23} = -\frac{3 \sin (2 \theta ) \left(\cos (2 \theta ) \left(-a_2+b_4+\lambda \right)+b_4-\lambda \right)}{2 \sqrt{2}}, \nonumber\\
&& \mathcal{S}_{26} = -\frac{1}{2} a_2 c_{\theta }^2-\lambda  s_{\theta }^2, \nonumber \\
&& \mathcal{S}_{27} = -\frac{a_2 c_{\theta }^2+2 \lambda  s_{\theta }^2}{\sqrt{2}}, \nonumber \\
&& \mathcal{S}_{33} = \frac{1}{4} \left(3 \cos (4 \theta ) \left(-a_2+b_4+\lambda \right)-a_2-3 b_4-3 \lambda \right), \nonumber \\
&& \mathcal{S}_{36} = \frac{\left(2 \lambda -a_2\right) c_{\theta } s_{\theta }}{\sqrt{2}}, \nonumber \\
&& \mathcal{S}_{37} = \left(2 \lambda -a_2\right) c_{\theta } s_{\theta }, \nonumber \\
&& \mathcal{S}_{44} = -a_2 s_{\theta }^2-2 \lambda  c_{\theta }^2, \nonumber \\
&& \mathcal{S}_{45} = \left(2 \lambda -a_2\right) c_{\theta } s_{\theta }, \nonumber \\
&& \mathcal{S}_{55} = -a_2 c_{\theta }^2-2 \lambda  s_{\theta }^2, \nonumber \\
&& \mathcal{S}_{66} = -3 \lambda, \nonumber \\
&& \mathcal{S}_{67} = -\sqrt{2} \lambda, \nonumber \\
&& \mathcal{S}_{77} = -4 \lambda.
\end{eqnarray}
For charge-1 channels, we have:
\[
\mathcal{S}_1 =
  \left[
  \begin{array}{ccc}
     -2 \lambda  c_{\theta }^2-a_2 s_{\theta }^2 & \left(2 \lambda -a_2\right) c_{\theta } s_{\theta } & 0
        \\
	 \left(2 \lambda -a_2\right) c_{\theta } s_{\theta } & -a_2 c_{\theta }^2-2 \lambda  s_{\theta }^2 & 0
	    \\
	     0 & 0 & -2 \lambda  \\
  \end{array}
  \right] .
\]
For the charge-2 channel with only one process, the matrix is simply given by $\mathcal{S}_2 = (-2 \lambda)$.

\section{\label{sec:tadpole}Connection with Potential where $v_s=0$}

The potential in Eq.~\ref{eq:v} can be written into a different form by translating the coordinate system 
of $(H,S)$ such that the EW vacuum has $\langle S \rangle =0$ (see e.g.,~\cite{Lewis:2017dme}). In this basis, there will generally be an additional 
tadpole term ($b_1 S$).
Making this translation of field variables leads to the same potential being represented with different potential
parameters, without changing the physics~\cite{Espinosa:2011ax}. So the scalar couplings as well as their masses and mixing angles wont
be affected by this translation. For easy comparison between these two representations, we show here the 
transformation rules between these two bases. Given potential parameters in the non-tadpole basis in Eq.~\ref{eq:v}, 
the parameters in the basis where $b_1 \neq 0$(denoted with a prime) can be obtained:
\begin{eqnarray}
&& b_1^{\prime} = v_s (b_2 + v_s(b_3 + b_4 v_s)), \nonumber \\
&& b_2^{\prime} = b_2 + v_s (2 b_3 +3 b_4 v_s), \nonumber \\
&& b_3^{\prime} = b_3 + 3 b_4 v_s, \nonumber \\
&& \mu^{2 \prime} = \mu^2 - \frac{1}{2} v_s (a_1 + a_2 v_s), \nonumber \\
&& a_1^{\prime} = a_1 + 2 a_2 v_s,
\end{eqnarray}
while $a_2, \lambda, b_4$ remains unchanged. On the other hand, given parameters in the tadpole basis 
where $v_s=0$ and $b_1 \neq 0$, the parameter set in the basis used in this work can be found:
\begin{eqnarray}
&& v_s = x, \nonumber \\
&& b_2 = b_2^{\prime} - x (2 b_3^{\prime} - 3 b_4^{\prime} x), \nonumber \\
&& b_3 = b_3^{\prime} - 3 b_4^{\prime} x, \nonumber \\
&& \mu^{2} = \mu^{2 \prime} + \frac{1}{2} x (a_1^{\prime} - a_2^{\prime} x), \nonumber \\
&& a_1 = a_1^{\prime} - 2 a_2^{\prime} x,
\end{eqnarray}
where $x$ is to be solved from the cubic equation 
\begin{eqnarray}
b_1^{\prime} - b_2^{\prime} x + b_3^{\prime} x^2 - b_4^{\prime} x^3=0,
\end{eqnarray}
which might give more than one solutions.
In the basis $v_s=0$, the degree of freedom carried by $v_s$ in the basis $v_s \neq 0$  is transformed to a 
different parameter. For example, one can choose it to be $a_2$ and then the full set of independent parameters can
be chosen as
\begin{eqnarray}
\centering
a_2, \quad \quad \mhb, \quad \quad \theta, \quad \quad b_3, \quad \quad b_4 . \quad
\end{eqnarray}
We note further there are also studies of this model where a $Z_2$ symmetry in the $S$ fields are imposed and are spontaneously 
broken~\cite{Pruna:2013bma,Robens:2015gla,Carena:2018vpt}. This specific model correspond to a special limit of the potential here. 

%\bibliographystyle{utphys}
%\bibliography{mybib}

\end{document}